\documentclass[AMA,STIX1COL]{WileyNJD-v2}
\usepackage{amsfonts}
\usepackage{amsmath}
\usepackage{xspace}
\usepackage{xcolor}
\usepackage{soul}
\usepackage{xspace}
\usepackage{url}
\usepackage{algorithm}
\usepackage{algpseudocode}
\usepackage{comment}
\usepackage{framed}
\usepackage{longtable}
\usepackage{tabularx}
\usepackage{ragged2e}
\usepackage{alltt}

\newcommand\BibTeX{{\rmfamily B\kern-.05em \textsc{i\kern-.025em b}\kern-.08em
T\kern-.1667em\lower.7ex\hbox{E}\kern-.125emX}}

\articletype{Article Type}%

\received{<day> <Month>, <year>}
\revised{<day> <Month>, <year>}
\accepted{<day> <Month>, <year>}

\excludecomment{1}

\usepackage{alltt}

\catcode`\|=\active
\def|{\mid}

\def\@p#1{\mathrel{\ooalign{\hfil$\mapstochar\mkern
      5mu$\hfil\cr$#1$}}}
\def \pfun      {\@p\fun}
\let \fun       \rightarrow

\def\comp{\mathrel{\raise 0.66ex\hbox{\oalign{\hfil%
        $\scriptscriptstyle\mathsf{o}$\hfil%
        \cr\hfil$\scriptscriptstyle\mathsf{9}$\hfil}}}}

\newcommand{\fp}{\tt} 

\newcommand{\true}{\mathbin{\mathsf{true}}}
\newcommand{\false}{\mathbin{\mathsf{false}}}

\renewcommand{\Cup}{\mathbin{\mathsf{un}}}
\newcommand{\Ncup}{\mathbin{\mathsf{nun}}}

\newcommand{\Disj}{\mathbin{\mathsf{disj}}}
\newcommand{\Ndisj}{\mathbin{\mathsf{ndisj}}}
\renewcommand{\Cap}{\mathbin{\mathsf{inters}}}
\newcommand{\Ncap}{\mathbin{\mathsf{ninters}}}
\newcommand{\Diff}{\mathbin{\mathsf{diff}}}
\newcommand{\Ndiff}{\mathbin{\mathsf{ndiff}}}
\newcommand*{\Size}{\mathbin{\mathsf{size}}}
\newcommand*{\Comp}{\mathbin{\mathsf{comp}}}
\newcommand{\Id}{\mathit{id}}
\newcommand{\Inv}{\mathit{inv}}
\newcommand{\Mod}{\mathbin{\mathsf{mod}}}

\newcommand{\jsetl}{\mbox{JSetL}\xspace}
\newcommand{\setlog}{$\{log\}$\xspace}
\newcommand{\CLPSET}{CLP($\mathcal{SET}$)\xspace}
\newcommand{\CLPFD}{CLP($\mathcal{FD}$)\xspace}
\newcommand{\CLPRIS}{$\LRIS$\xspace}

\newcommand{\SATRIS}{\mathit{SAT}_\mathcal{RIS}}

\newcommand{\LRIS}{\mathcal{L}_\mathcal{RIS}}

\newcommand{\LX}{\mathcal{L}_\mathcal{X}}
\newcommand{\LSETFD}{\mathcal{L}_\mathcal{SET}}

\newcommand{\plus}{/}
\newcommand{\ww}{\{\cdot \plus \cdot\}}
\newcommand{\e}{\emptyset}
\newcommand{\disj}{\parallel}

\newcommand{\ris}[4]{\{ #1 :\, #2\,|\,#3\,\mathbin{\bullet}\,#4\}}
\newcommand{\riss}[3]{\{ #1\,|\,#2\,\mathbin{\bullet}\,#3\}}
\newcommand{\risnopattern}[3]{\{ #1 :\, #2\,|\,#3\}}

\newcommand{\Var}{\mathcal{V}}
\newcommand{\Ur}{\mathcal{X}}

\newcommand{\ct}{c}       
\newcommand{\flt}{\mathcal{F}}   
\newcommand{\ptt}{p}      

\newcommand{\type}{\mathcal{T}}
\newcommand{\stype}{\mathsf{S}}
\newcommand{\ptype}{\mathsf{P}}
\newcommand{\utype}{\mathsf{U}}

\newcommand{\why}[1]{\tag*{{\footnotesize [by #1]}}}

\renewcommand{\iff}{\Leftrightarrow}

\definecolor{formula}{gray}{0.9}

\newtheorem{example}{Example}

\begin{document}

\title{Declarative Programming with Intensional Sets in Java Using JSetL}

\author[1]{Maximiliano Cristi\'a*}

\author[2]{Gianfranco Rossi}

\author[2]{Andrea Fois}

\authormark{MAXIMILIANO CRISTI\'A \textsc{et al}}

\address[1]{\orgname{Universidad Nacional de Rosario and CIFASIS}, \country{Rosario, Argentina}}

\address[2]{\orgname{Universit\`a di Parma}, \country{Parma, Italy}}

\corres{*Maximiliano Cristi\'a, Tucum\'an 4142 -- (2000) Rosario -- Argentina. \email{cristia@cifasis-conicet.gov.ar}}

\abstract[Abstract]{%
Intensional sets are sets given by a property rather than by enumerating their
elements. In previous work, we have proposed a decision procedure for a
first-order logic language which provides Restricted Intensional Sets (RIS),
i.e., a sub-class of intensional sets that are guaranteed to denote
finite---though unbounded---sets. In this paper we show how RIS can be
exploited as a convenient programming tool also in a conventional setting,
namely, the imperative O-O language Java. We do this by considering a Java
library, called JSetL, that integrates the notions of logical variable, (set)
unification and constraints that are typical of constraint logic programming
languages into the Java language. We show how JSetL is naturally extended to
accommodate for RIS and RIS constraints, and how this extension can be
exploited, on the one hand, to support a more declarative style of programming
and, on the other hand, to effectively enhance the expressive power of the
constraint language provided by the library.
}

\keywords{Java; JSetL; declarative programming; set programming; logic programming; constraint programming; set theory; set constraint}

\maketitle

\catcode`\@=\active
\def@{\mathbin{\bullet}}

\section{\label{intro}Introduction and motivations}

\emph{Set-oriented programming} is a programming paradigm where the well-known
mathematical notion of set plays a fundamental role in providing high-level
(declarative) descriptions of problem solutions. This approach is well
exemplified by specification languages such as Z \cite{Woodcock00} and B
\cite{schneider2001b} and programming languages such as SETL
\cite{DBLP:books/daglib/0067831}. Set-oriented programming is also supported to
some extent by other general-purpose programming languages, such as Claire
\cite{DBLP:journals/tplp/CaseauJL02}, Miranda \cite{DBLP:journals/eatcs/Turner87} and Bandicoot \cite{bandicoot}.
The main goal of set-oriented programming is to support rapid
software prototyping. However, a set-oriented approach can be of
great help also in other software development activities, such as
program verification, since it provides a valuable tool for the
development of correct-by-construction prototypes.

In the nineties many proposals for set-oriented programming have emerged in the
field of \emph{declarative programming}, where sets and operations on sets are
added to a first-order logic language as first-class entities of the
language
\cite{Abiteboul1991,Jayaraman1995,Liu1998,DBLP:journals/jlp/DovierOPR96,DBLP:journals/constraints/Gervet97,Dovier00}.
Efforts in this direction are well attested by the two workshops on
Logic/Declarative Programming with Sets held in the nineties
\cite{DBLP:conf/iclp/1993w5,workshop1999}. Among these proposals, \CLPSET
\cite{Dovier00} is particularly linked to our work. In effect, \CLPSET is a
\emph{constraint logic programming} (CLP) language whose constraint domain is
that of \emph{hereditarily finite sets}, i.e., finitely nested sets that are
finite at each level of nesting. \CLPSET\ allows the user to operate with
partially specified sets through a number of primitive constraints representing
all the most commonly used set-theoretic operations (e.g., union, intersection,
difference). A complete constraint solver for this language is provided,
capable of deciding the satisfiability of arbitrary conjunctions of primitive
constraints.

These ideas and results have been implemented also in the context of more
conventional programming languages, namely Java, in the form of the JSetL
library \cite{Rossi2007}. JSetL implements the notions of logical variable, (set)
unification and constraints, that are typical of constraint logic programming
languages, into the Java language. JSetL supports \emph{declarative constraint
programming with sets}. This style of programming is illustrated
\cite{Rossi2007,Bergenti2011,DBLP:journals/fuin/RossiB15} by a number of sample programs showing how JSetL
facilities, such as partially specified sets, set unification, non-determinism,
can be of great help in devising and implementing high-level declarative
solutions for many, possibly complex, problems.

In the practice of mathematics, however, it is common to distinguish between
sets designated via explicit enumeration (\emph{extensional sets})---e.g.,
$\{a,b,c\}$\/---and sets described through the use of properties and/or
characteristic functions (\emph{intensional sets})---e.g., $\{x\: : \:
\varphi(x)\}$\/. Intensional sets are widely recognized as a key feature to
describe complex problems. Having the possibility to represent and manipulate
intensional sets could constitute a valuable improvement in the expressive
power of a programming language. Notwithstanding, very few programming
languages provide support for intensional sets---e.g., SETL
\cite{DBLP:books/daglib/0067831} and Python. In these proposals, however, set
operators are applied only to intensional sets that denote completely specified
sets, i.e., sets where all elements have a known value. Moreover, often these
proposals are mostly concerned with aggregate operations (e.g., finding the
maximum or the minimum or the sum of all elements of the set), and place
limited attention to other basic set-theoretical facilities.
Ultimately, in all these proposals the language does not provide real direct
support for \emph{reasoning} about intensional sets. Conversely,
languages/libraries supporting declarative programming with sets, such as
\CLPSET\ and JSetL, provide general set-theoretic operations on
\emph{extensional sets}, usually in the form of \emph{(set) constraints}, that
allow set objects to be manipulated even if they are represented by variables
or they are only partially specified.

In previous works \cite{DBLP:conf/iclp/DovierPR03,DBLP:conf/cade/CristiaR17},
we have extended a first-order logic language providing extensional sets to
support also intensional sets. This is done by introducing them as first-class
entities of the language and providing operations on them as primitive
constraints (i.e., \emph{intensional set constraints}). The proposed constraint
solver is able to deal with intensional sets \emph{without} explicitly
enumerating all their elements. In particular, \emph{Restricted Intensional
Sets} (RIS) \cite{DBLP:conf/cade/CristiaR17} have proved to be much more
effective from a programming viewpoint than general intensional sets.

RIS have similar syntax and semantics to the set comprehensions available in
the Z formal specification language, i.e. $\{ \ct : D | \flt @ \ptt(\ct) \}$,
where $D$ is a finite set, $\flt$ is a quantifier-free formula over a
first-order theory $\Ur$, and $\ct$ and $\ptt$ are $\Ur$-terms. Intuitively,
the semantics of $\{ \ct : D | \flt @ \ptt(\ct) \}$ is ``the set of terms
$\ptt(\ct)$ such that $\ct$ is in $D$ and $\flt$ holds''. We say that this
class of intensional sets is \emph{restricted} because they denote
\emph{finite} sets, while in Z they can be infinite. The finiteness of $D$,
along with a few restrictions on variables occurring in $\flt$ and $\ptt$,
guarantees that the RIS is a finite set, given that its cardinality is as large
as D's. Nonetheless, RIS can be not completely specified. In particular, as the
domain can be a variable, RIS are finite but \emph{unbounded}.

The initial work on RIS \cite{DBLP:conf/cade/CristiaR17}, however, is mainly concerned with
the definition of the constraint (logic) language---called $\LRIS$---and its
solver, and the proof of soundness and completeness of the constraint solving
procedure. In this paper, our main aim is to explore \emph{programming} with
(restricted) intensional sets. Specifically, we are interested in exploring the
potential of using RIS in the more conventional setting of imperative O-O
languages. To this purpose, we consider the Java library JSetL. First, we show
how JSetL is naturally extended to accommodate for RIS. Then, we show with a
number of simple examples how this extension can be exploited, on the one hand,
to support a more declarative style of programming and, on the other hand, to
effectively enhance the expressive power of the constraint language provided by
the library. It is worth noting that although we are focusing on Java, the same
considerations can be easily ported to other O-O languages, such as C++.

Our claim is that the language of RIS constraints is expressive enough to allow
usual programming solutions to be encoded as formulas in that language. In
particular, the fact that RIS can also be \emph{recursively} defined, together
with the fact that ordered pairs can be set elements, makes it possible to
define most of the classic recursive functions in a set-oriented fashion. That
is, in our language, functions are sets of ordered pairs that are managed by
means of classic set theoretic operators such as equality and membership.

The paper is organized as follows. Section \ref{theory} introduces the
theoretical framework underlying RIS. Section \ref{jsetl} briefly reviews the
JSetL library, while Section \ref{jsetl-ris} presents the extension of JSetL
with RIS. In Section \ref{declarative} we start showing examples using JSetL to
demonstrate the usefulness of RIS and RIS constraints to support declarative
programming; in particular, Section \ref{pfun} shows how RIS can be used to
define and manipulate partial functions. In Section \ref{extensions} we
consider some extensions to RIS and we present examples showing their
usefulness. In Section \ref{implementation} we briefly address some design and
implementation issues. Section \ref{practice} provides a
quantitative evaluation and some general considerations on the practical
usability of RIS and set-oriented programming, in general. A comparison of our
approach with related work and our conclusions are presented in Sections
\ref{related} and \ref{conclusion}, respectively.


\section{A Theory of RIS}\label{theory}

In this section we introduce the theoretical framework underlying the JSetL
library, with special reference to the support it offers to RIS.

The language that embodies RIS, called $\LRIS$, is a quantifier-free
first-order logic language which provides both RIS and extensional sets, along
with basic operations on them, as primitive entities of the language. $\LRIS$
is \emph{parametric} with respect to an arbitrary theory $\Ur$, for which a
decision procedure for any admissible $\Ur$-formula is assumed to be available.
Elements of $\LRIS$ sets are the objects provided by $\Ur$, which can be
manipulated through the primitive operators that $\Ur$ offers. The $\Ur$
language, called $\LX$, is assumed to provide at least, equality ($=_\Ur$) and
inequality ($\neq_\Ur$), a not empty collection of constants, $a_1$, $a_2$,
$\dots$, and a binary function symbol to represent ordered pairs, e.g.,
$(a_1,a_2)$.

Besides, the function and predicate symbols provided by $\Ur$, $\LRIS$ provides
special set constructors and a handful of reserved predicate symbols endowed
with a pre-designated set-theoretic meaning. Set constructors are used to build
set terms.

\begin{definition}[Set terms]
A \emph{set term} is any $\LRIS$ term of one of the following forms:
\begin{enumerate}
\item[$-$] $\emptyset$ (\emph{empty set});
\item[$-$] $\{x \plus A\}$ (\emph{extensional set term}),  where $x$,
called \emph{element part}, is an $\Ur$-term, and $A$, called \emph{set part},
is a set term;
\item[$-$] $\{ \ct : D | \flt @ \ptt(\ct) \}$ (\emph{RIS term}),
where $\ct$, called \emph{control term}, is an $\Ur$-term; $D$, called
\emph{domain}, is a set term; $\flt$, called \emph{filter}, is an
$\Ur$-formula; and $\ptt$, called \emph{pattern}, is an $\Ur$-term containing
$\ct$;
\item[$-$] any variable belonging to a denumerable set of variables $\Var_\mathcal{S}$
(\emph{set variables}).
 \qed
\end{enumerate}
\end{definition}

Intuitively, an extensional set term $\{x \plus A\}$ is interpreted as $\{x\}
\cup A$. A RIS term is interpreted as follows:
if $x_1,\dots,x_n$ ($n > 0$) are all the variables occurring in $\ct$, then:
$$\{ \ct : D | \flt @ \ptt(\ct) \}$$
denotes the set:
$$\{y | \exists x_1,\dots,x_n (\ct \in D \land \flt \land y =_\Ur \ptt(\ct))\}$$
where $x_1,\dots,x_n$ are bound variables whose scope is the RIS term
itself. Hence,
$\LRIS$ set terms represent \emph{untyped unbounded finite hybrid sets}, i.e.,
unbounded finite sets whose elements are of arbitrary sorts.
%

\begin{remark}[Notation] \label{notation}
As a notational convenience, $\{t_1 \plus \{t_2 \plus \cdots \{ t_n \plus
t\}\cdots\}\}$ (resp., $\{t_1 \plus \{t_2 \plus \cdots \{ t_n \plus
\e\}\cdots\}\}$), $n \ge 1$,
is written as $\{t_1,t_2,\dots,t_n \plus t\}$ (resp., $\{t_1,t_2,\dots,t_n\}$).
When useful, the domain of a RIS can be represented also as an interval
$[m,n]$, $m$ and $n$ integer constants, which is intended as a shorthand for
$\{m,m+1,\dots,n\}$. When the pattern is the same as the control term, the
former can be omitted (as in Z). Furthermore, the following name conventions
will be used throughout the paper: $A,B,C,D,E,S$ stand for arbitrary sets
(either variable or not), while $X,Y,Z,N$ stand for variable sets; $R$ stands
for a RIS; and $x,y,z,n$ are variables representing set elements.
\end{remark}

It is important to observe that elements and sets in both extensional set terms
and RIS terms can be variables.

\begin{definition}
If $s$ is a set term, we say that $s$ denotes a \emph{partially specified set}
if either $s$ is a variable; or $s$ is $\{t_1,\dots,t_n\}$ and at least one
$t_i$ is a variable or a term containing a variable; or $s$ is $\{t_1,\dots,t_n
\plus t\}$ and $t$ is a variable; or $s$ is a RIS term and its domain is a set
term denoting a partially specified set.
\end{definition}

The following are simple examples of RIS terms.

\begin{example} \label{ex:first}
Assume that $\LX$ provides the constant, function and predicate symbols of the
theory of the integer numbers. Let $x$, $y$ and $z$ be $\Ur$-variables (i.e.,
variables ranging over the domain of $\Ur$) and let $D$ and $S$ be set
variables. The following are RIS terms:
\begin{enumerate}
\item[$i.$] $\{x:[-2,2] | x\ \Mod\ 2 = 0 @ x\}$ (also written as
$\{x:[-2,2] | x\ \Mod\ 2 =
0\}$)
\item[$ii.$] $\{x:D | x > 0 @ (x,x*x)\}$, where $D$ is a free variable in the RIS
\item[$iii.$] $\{(x,y):\{z \plus A\} | y \neq 0 @ (x,y)\}$, where $z$ and $A$ are
free variables.   \qed
\end{enumerate}
\end{example}

\begin{definition}[$\LRIS$ constraints]
A \emph{(primitive) $\LRIS$ constraint} is any $\LRIS$ atom of one of the
following forms: $A = B$, $A \neq B$, $e \in A$, $e \notin A$, $\Cup(A,B,C)$
and $\Disj(A,B)$, where $A$, $B$ and $C$ are $\LRIS$ set terms and $e$ is an
$\Ur$-term.
\end{definition}

The intuitive meaning of the $\LRIS$ constraints is: $A = B$ (resp., $A \neq
B$) represents set equality (resp., inequality) between the sets denoted by $A$
and $B$; $e \in A$ (resp., $e \notin A$) represents set membership (resp., not
membership); $\Cup(A,B,C)$ represents set union, i.e., $C = A \cup B$; and
$\Disj(A,B)$ represents set disjunction, i.e., $A \cap B = \emptyset$
(also denoted as $A \disj B$).

\emph{$\LRIS$ formulas} are built from $\LRIS$ constraints using conjunction
and disjunction in the usual way.

The collection of predicate symbols used for the primitive constraints turns
out to be sufficient to define constraints implementing other common set
operators \cite{Dovier00}. Specifically, the following atoms are provided by
$\LRIS$ as defined formulas: $A \subseteq B$ (interpreted as set inclusion),
$\Cap(A,B,C)$ (interpreted as $C = A \cap B$), $\Diff(A,B,C)$ (interpreted as
$C = A \backslash B$). As an example, $A \subseteq B$ is defined by the $\LRIS$
formula $\Cup(A,B,B)$. We will refer to these atoms as \emph{derived
constraints}. Whenever a formula contains a derived constraint, the constraint
is replaced by its definition turning the given formula into a $\LRIS$ formula.

The same approach is used to introduce the negative counterparts of  set
operators not defined as primitive constraints. Specifically, derived
constraints are introduced for $\lnot\cup$ and $\lnot\disj$ (called $\Ncup$ and
$\Ndisj$, respectively), as well as for $\lnot\subseteq$, $\lnot\cap$ and
$\lnot\backslash$ (called $\not\subseteq$, $\Ncap$ and $\Ndiff$, respectively).
Observe that, thanks to the availability of the negated versions of set
operators as derived constraints (hence, as positive atoms), classical negation
is not necessary in $\LRIS$.

$\LRIS$ provides a complete \emph{constraint solver}, for a large fragment of
its input language. This solver is able to decide the satisfiability of any
admissible $\LRIS$ formula. Intuitively, non-admissible formulas are those
where a variable $A$ is the domain of a RIS representing a function and, at the
same, time $A$ is either a sub or a superset of that function. For example, the
$\LRIS$ formula $\riss{x : D}{\true}{(x,y)} \subseteq D \land D \neq \e$ is
non-admissible since it implies that if $z \in D$ then $(z,n_1) \in D$ and so
$((z,n_1),n_2) \in D$ ($n_i$ fresh variables) and so forth, thus generating an
infinite $\Ur$-term.
In the rest of this paper, we will restrict our attention to admissible $\LRIS$
formulas.

The $\LRIS$ constraint solver reduces any input formula $\Phi$ to either
$\false$ (hence, $\Phi$ is unsatisfiable), or to an equi-satisfiable
disjunction of formulas in a simplified form, called the \emph{solved form},
which is guaranteed to be satisfiable (hence, $\Phi$ is satisfiable). If $\Phi$
is satisfiable, the answer computed by the solver constitutes a finite
representation of all the concrete (or ground) solutions of the input formula.
See \ref{app:formal} for a formal definition of admissible and
solved form $\LRIS$ formula.

The following examples show simple $\mathcal{RIS}$-formulas involving RIS and
their processing by the $\LRIS$ constraint solver.

\begin{example}[$\LRIS$ constraint solving]\label{ex:solving}
\mbox{}
\begin{enumerate}
\item[$i.$]
The $\LRIS$ constraint $(5,y) \in \{x:D | x > 0 @ (x,x*x)\}$ is rewritten by
the solver to the solved form formula $y = 25 \land D = \{5 \plus N_1\}$, where
the second equality states that $D$ must contain $5$ and something else,
denoted $N_1$.
\item[$ii.$]
The formula $S = \{2,4,6\}\; \land\; S = \{x : D | x\:\:\Mod\:\:2 = 0\}$
is rewritten by the solver to a solved form formula containing the
constraint $D = \{2,4,6 \plus N_1\} \land  \{x : N_1 | x\: \Mod \: 2
= 0\}=\e$, where the second equality states that $N_1$ cannot
contain even numbers (note that this constraint has the obvious
solution $N_1 = \e$).
\item[$iii.$] The formula $A = \{x:D | x \neq 0\} \land \Cup(A,B,C) \land \Disj(A,C) \land A \neq
\emptyset$
is rewritten by the solver to $\false$ (as a matter of fact, $\Cup(A,B,C) \land
\Disj(A,C)$ is satisfiable only if $A = \emptyset$); hence, the formula is
unsatisfiable.
 \qed
\end{enumerate}
\end{example}

In order to allow the solver to act as a decision procedure for a large part of
its input language, the control term $\ct$ and the pattern $\ptt$ of a RIS are
restricted to be of specific forms. All RIS shown above, meet these
restrictions.

\begin{definition}[Admissible control terms and patterns]\label{def:admissible}
If $x$ and $y$ are $\Ur$-variables, then an \emph{admissible control term}
$\ct$ is either $x$ or $(x,y)$, while an \emph{admissible pattern} $\ptt$ is
either $\ct$ or $(\ct,s)$, where $s$ is any $\Ur$-term, possibly involving the
variables in $\ct$. \qed
\end{definition}

As it will be evident in Section \ref{extensions}, these
restrictions could be often relaxed in practice.

\subsection{An instance of $\LRIS$}

$\LRIS$ is \emph{parametric} with respect to the theory $\Ur$. In the rest of
this paper we will consider a specific instance of $\LRIS$, indicated as
$\LRIS(\mathcal{SET})$, where $\Ur$ is the theory $\mathcal{SET}$.

$\mathcal{SET}$ is basically the theory of \emph{hereditarily finite hybrid
sets} \cite{Dovier00}, augmented with the theory underlying \CLPFD,
that is integer arithmetic over finite domains
\cite{Palu:2003:IFD:888251.888272}. The constraint language of this theory,
here simply called $\LSETFD$, provides the same function symbols as $\LRIS$ for
building extensional set terms (namely, $\e$ and $\ww$), along with a
collection of predicate symbols including those of $\LRIS$, with the same
interpretation. In addition, $\LSETFD$ provides the usual function symbols
representing operations over integer numbers (e.g., $+,-,\Mod$, etc.), as well
as the predicate symbols $\Size$, representing set cardinality, and $\leq$,
representing the order relation on the integers. One notable difference w.r.t.
$\LRIS$ is that set elements can be either finite sets or non-set elements of
any sort (i.e., nested sets are allowed).

The theory $\mathcal{SET}$ is endowed with a constraint solver
that combines the (set) constraint solving technique by Dovier et al. \cite{Dovier00} with
those of CLP(FD) \cite{Codognet00}, namely integer constraint solving over finite
domains. The constraint solver for $\mathcal{SET}$ is proved to be a decision
procedure for its formulas, provided a finite domain is associated to each
integer variable occurring in the input formula.

\begin{example}[$\LRIS(\mathcal{SET})$ formulas] \label{ex:clpset}
The following formula written in $\LRIS(\mathcal{SET})$  states the equality
between an extensional set and a RIS computing all the subsets of cardinality 2
of a given set:
\[
 \{X:\{\{1,3\},2,\{1\}\} | \Size(X,2)\} = \{\{1,3\}\}
\]
This formula is (correctly) proved by the $\LRIS(\mathcal{SET})$ solver to be
$\true$. \qed
\end{example}

Note that we are using the same external notation for both $\LRIS$ set terms
and $\LSETFD$ set terms. However, which kind of set terms we are actually
referring to is automatically inferred from the context where the terms occur.


\section{An Informal Introduction to JSetL}\label{jsetl}

In this section, we introduce JSetL, a Java library that supports declarative
(constraint) programming in an O-O framework \cite{Rossi2007}. To this end, JSetL
combines the object-oriented programming paradigm of Java with valuable
concepts of CLP languages, such as logical variables, unification, constraint
solving and non-determinism. Specifically, JSetL implements $\LSETFD$ in Java.
As such it provides, among others, very general forms of (possibly, partially
specified) extensional sets, along with most of the usual set-theoretical
operators (e.g., set equality, membership, union, inclusion, etc.) as
constraints.

JSetL has many similarities with proposals for libraries supporting
constraint programming, in particular those providing set variables and set
constrains, such as Choco \cite{choco}, FaCiLe \cite{brisset:hal-00938018}, JaCoP
\cite{jacop}, Gecode \cite{gecode}. As evidence of this, JSetL has been used
also as one of the first six implementations for the standard Java Constraint
Programming API defined in the Java Specification Request JSR-331 and an
implementation based on the new version of JSetL is included in the latest
release of JSR-331 made available through public GitHub and MVN repositories.
The main distinguishing feature of JSetL w.r.t. other libraries for constraint
programming is the availability of partially specified sets \cite{Bergenti2011}
(including set unification \cite{Dovier2006} to deal with them), and the support it
offers to users for nondeterministic programming \cite{DBLP:journals/fuin/RossiB15}.

The JSetL library can be downloaded from the JSetL's home page at
\url{http://www.clpset.unipr.it/jsetl/}.

The fundamental data abstraction to support declarative constraint programming
in JSetL is that of \emph{logical object}. Basically, logical objects occur in
the form of logical variables and logical collections.

\emph{Logical variables} represent ``unknowns''. Differently from ordinary
programming language variables, logical variables have no modifiable value
stored in them. In fact, values are associated to logical variables through
constraints, representing relations over some specific domains.
In JSetL, logical variables are instances of the class {\fp LVar}. When
created, an {\fp LVar} object can be either \emph{uninitialized} (i.e., its
value is unknown) or \emph{initialized} (i.e., its value is bound to some
specific value). Moreover, each {\fp LVar} object can have an optional external
name (namely, a string) which can be useful when printing the variable and the
possible constraints involving it (see Example \ref{ex:lvar}).
{\fp LVar} objects can be manipulated through constraints, namely equality
({\fp eq}), inequality ({\fp neq}), set membership ({\fp in}) and not
membership ({\fp nin}). The library provides also utility methods to test
whether a variable is initialized or not, to get the value of a initialized
variable (but not to modify it), to get/set its external name, and so on.

\begin{example}[Logical variables] \label{ex:lvar}
\mbox{}

\begin{alltt}
   LVar x = new LVar();      // an uninitialized logical variable
   LVar y = new LVar("y",1); // an initialized logical variable
                             // with external name "y" and value 1
   x.setName("x");           // set the external name of x to "x"
   y.output();               // print the value bound to y
\end{alltt}
 \noindent Executing {\fp y.output()} will print on the standard output
{\fp \_y = 1}, i.e., the external name of the logical variable followed by its
value (or {\fp unknown} if the variable is uninitialized).\footnote{The
printed external name of a logical object is prefixed by the character {\fp
'\_'} in order to better distinguish variable names from ordinary strings.}
\qed
\end{example}

Values associated with generic logical variables can be of any type. For some
specific domains, however, JSetL offers specializations of the {\fp LVar} data
type, which provide further specific constraints. In particular, for the domain
of integers, JSetL offers the class {\fp IntLVar}, which extends {\fp LVar}
with a number of new methods and constraints specific for integers. An
\emph{integer logical expression} is created using methods implementing
arithmetic operations, such as {\fp mul}, {\fp mod}, etc., applied to {\fp
IntLVar} objects, and returning {\fp IntLVar} objects. Moreover, {\fp IntLVar}
provides integer comparison constraints, such as $<$ ({\fp lt}), $\leq$ ({\fp
le}), etc. to relate integer logical expressions to each other.

Another important kind of logical objects are logical collections, namely,
\emph{logical sets} and \emph{lists}. Hereafter we will focus on logical sets,
but most of the following considerations apply also to logical lists. The value
of a logical set is a collection of elements of any type, including other
logical objects. Logical sets can be \emph{partially specified} \cite{Bergenti2011},
in that they can contain uninitialized logical objects as elements, as well as
an uninitialized logical set as the rest of the set.

In JSetL, logical sets are instances of the class {\fp LSet}, which in turn is
a subclass of the class {\fp LCollection}. Values of {\fp LSet} objects are
instances of the class {\fp HashSet} which implements the {\fp java.util}
interface {\fp Set}.

\begin{remark}
For the sake of clarity, we will adopt here the same syntactic notations for
names used in $\LRIS$---see Remark \ref{notation}---although it may sometimes
conflict with the conventions usually adopted in Java.
\end{remark}

\begin{example} [Logical sets] \label{ex:jsetl_lsets}
\mbox{}

\begin{alltt}
   LSet S1 = new LSet("S1"); // an uninitialized logical set
                             // with external name "S1"
   LSet S2 = LSet.empty().ins(1,2);    // the set \{1,2\}
   LVar x = new LVar("x");
   LSet S3 = S1.ins(x);      // the set \{x\} \(\cup\) S1      \qed
\end{alltt}
\end{example}

{\fp empty()} is a static method of the class {\fp LSet} returning the
empty set. {\fp ins} is the \emph{element insertion} method for {\fp LSet}
objects: {\fp S.ins(o$_1$,\dots,o$_n$), $n \ge 1$,} returns the new logical set
whose elements are those of the set ${\fp S} \cup \{{\fp o}_1,\dots,{\fp
o}_n\}$. In particular, the last statement in Example \ref{ex:jsetl_lsets}
creates a partially specified set {\fp S3} containing an unknown element {\fp
x} and an unknown rest {\fp S1} (i.e., {\fp $\{$x$\:\plus\:$S1$\}$}, using the
abstract notation of Remark \ref{notation}).

A number of constraints are provided to work with {\fp LSet} objects, which
extend those provided by {\fp LVar}. In particular, {\fp LSet} provides
equality and inequality constraints that account for the semantic properties of
sets (namely, irrelevance of order and duplication of elements); moreover it
provides constraints for many of the standard set-theoretical operations, such
as union ({\fp union}), intersection ({\fp inters}), inclusion ({\fp subset}),
and so on.

Constraints are instances of the library class {\fp Constraint}. They are
solved using a \emph{constraint solver} that implements the $\LSETFD$ solver in
Java.

A constraint solver in \jsetl\ is an instance of the class {\fp Solver}.
Basically, it provides methods for adding constraints to its \emph{constraint
store} (e.g., the method {\fp add}) and to prove constraint satisfiability
(e.g., the method {\fp solve}). If {\fp solver} is a solver, $\Gamma$ is the
collection of constraints stored in its constraint store (possibly empty), and
{\fp C} is a constraint, then {\fp solver.solve(C)} checks whether $\Gamma\,
\wedge\, {\fp C}$ is satisfiable or not, i.e., whether there exists an
assignment of values to the logical variables of $\Gamma\, \wedge\, {\fp C}$
that makes this formula $\true$ in the intended interpretation; if
$\Gamma\, \wedge\, {\fp C}$ is unsatisfiable, then {\fp solve} raises the
exception {\fp Failure}. The order in which constraints are posted to the
solver is irrelevant.

The class {\fp Solver} implements the constraint solver of $\LRIS$.
Any conjunction of
constraints is reduced to a simplified form---the \emph{solved
form}---which is proved to be satisfiable. The success of this reduction allows
one to conclude the satisfiability of the original collection of constraints.
On the other hand, the detection of a failure (logically, the reduction to
$\false$) implies the unsatisfiability of the original constraint. Solved form
constraints are \emph{irreducible} constraints. As such, they are left in the
constraint store and possibly passed ahead to a new invocation of the
constraint solver. A successful computation, therefore, can terminate with a
collection of solved form constraints in the final constraint store.

\begin{example} [Constraint solving] \label{ex:jsetl_constraint_solving}
\mbox{}
\begin{alltt}
   LSet S1 = LSet.empty().ins(1,new LVar("z"));  // the set \{1,z\}
   LVar x = new LVar("x"), y = new LVar("y");
   LSet S2 = LSet.empty().ins(x,y);       // the set \{x,y\}
   Solver solver = new Solver();
   solver.add(S1.eq(S2).and(x.neq(1)));   // the constraint S1=S2 \(\land\) x\(\neq\)1
   solver.solve();
   y.output();
   solver.showStore();
\end{alltt}

\noindent The method {\fp showStore} prints all the constraints stored in the
constraint store. Executing this code fragment will output: {\fp \_y = 1}, {\fp
Store:~\_x neq 1}. \qed
\end{example}

The following is an example that illustrates the declarative programming style
supported by JSetL. It exploits the nondeterminism embedded in set operations.
Solving equalities, as well as other basic set-theoretical operations, over
partially specified sets yields, in general, multiple solutions. The JSetL
solver is able to nondeterministically compute all these solutions, by means of
backtracking. In this and in all other examples in the paper, {\fp solver}
represents an instance of the class {\fp Solver}.

\begin{example}[Permutations]\label{ex:permutations}
Print all permutations of an array $A$ of \emph{distinct} integer numbers. The
problem can be modelled as a \emph{set unification} problem \cite{Dovier2006}, where
the set $E$ of all elements of $A$ is unified with a (partially specified) set
of $\lvert E \rvert$ logical variables, i.e., $E = \{x_1,\dots,x_{\lvert E
\rvert}\}$. Each solution to this problem, that is, each assignment of values
to variables $x_1,\dots,x_{\lvert E \rvert}$, represents a possible permutation
of the integers in $A$. Note that set unification computes all such
permutations as
the order of set elements is immaterial to establish equality between two
sets.

\begin{alltt}
   public static void allPermutations(Integer[] A)\{
        LSet E = LSet.empty().insAll(A);
        LSet S = LSet.mkSet(A.length);
        solver.add(E.eq(S));
        solver.forEachSolution(i -> \{System.out.print(i+")  ");
                                     S.printElems(' ');\});
   \}
\end{alltt}

\noindent The method {\fp allPermutations} takes an array of integers {\fp A}
and calls the JSetL method {\fp printElems} for each permutation of the input
array. The first line creates a logical set {\fp E} out of the elements of
array {\fp A}. The second line creates a logical set {\fp S} which contains as
many logical variables as the length of {\fp A}. The third line adds the set
equality constraint ${\fp E} = {\fp S}$ to the constraint store. The solutions
of this constraint will non-deterministically assign all the values in {\fp A}
to the variables in {\fp S}. The last line asks the solver to execute the given
statements for each solution of the constraint added above. {\fp i} represents
the index of the computed solution; {\fp printElems} prints all elements of the
set {\fp S}, separated by the specified character. The following is an example
of usage of the method {\fp allPermutations}.

\begin{alltt}
   Integer[] elems = \{1, 2, 3\};
   allPermutations(elems);
\end{alltt}

\noindent The output produced by executing this code is:
\begin{alltt}
   1)  1 2 3
   2)  1 3 2
   3)  2 1 3
   4)  2 3 1
   5)  3 1 2
   6)  3 2 1 \normalsize{\qed}
\end{alltt}
\end{example}

Since its first release \cite{Rossi2007}, JSetL has evolved in various
respects \cite{jsetlman}. In particular, its constraint solving
capabilities have been enhanced by adding Finite Domain (FD) constraints, both
on integers (class {\fp IntLVar}), and on sets of integers (class {\fp
SetLVar}). Moreover, new data abstractions are provided, namely intervals,
multi-intervals, and set intervals, possibly to be used in conjunction with FD
constraints. Basically, JSetL implements
the original proposals for FD constraint solving  \cite{Codognet00} \cite{DBLP:journals/constraints/Gervet97},
integrating them within the \CLPSET constraint solver
\cite{Palu:2003:IFD:888251.888272}. More recently, JSetL has been also extended
with new classes to support the notions of (logical) binary relations (class
{\fp LRel}) and partial functions (class {\fp LMap})
\cite{DBLP:journals/jlp/DovierOPR96,DBLP:journals/jar/CristiaR20}. These
classes extends the class {\fp LSet}, and objects created out of them can be
manipulated through the usual set-theoretic constraints as well as through new
ad-hoc relational constraints (e.g., the constraint $\Comp$ for relation
composition).


\section{RIS in JSetL}\label{jsetl-ris}

In this section we show how JSetL can be naturally extended to implement
$\LRIS(\mathcal{SET})$, that is to provide RIS and constraints over RIS in
conjunction with the set abstractions which are already available in JSetL.

Actually, JSetL implements an extended version of the language of
RIS described in Section \ref{theory}. In this section, we will
focus on the simpler version of RIS (basically that presented in
Section \ref{theory}), while extended RIS will be discussed in
detail in Section \ref{extensions}.

\subsection{RIS Data Abstraction}\label{ssec:ris}

\begin{definition}[RIS]
A \emph{Restricted Intensional Set} (RIS) is an instance of the class {\fp
Ris}, created by the {\fp Ris} constructor:

\begin{alltt}
   Ris(LObject c,
       LSet D,
       Constraint F,
       LObject p,
       LObject... dummyVars)
\end{alltt}

\noindent where {\fp c} is the control term, {\fp D} the domain, {\fp F} the
filter, {\fp p} the pattern, and {\fp dummyVars} a possibly empty sequence of
logical variables (using the Java \emph{varargs} construct to pass an
arbitrary number of objects to the method). The pattern {\fp p} can be omitted
if it coincides with the control term {\fp c}, provided {\fp dummyVars} is
empty. \qed
\end{definition}

\begin{example}[{\fp Ris} object creation]\label{ex:Ris obj}
The RIS $\{x:[-2,2] | x\ \Mod\ 2 = 0 @ x\}$ (see Example
\ref{ex:first}) is created in JSetL as follows:
\begin{alltt}
   IntLVar x = new IntLVar();
   Ris R = new Ris(x,new IntLSet(-2,2),x.mod(2).eq(0));
\end{alltt}
\noindent where {\fp IntLSet(-2,2)} represents the closed (integer) interval
$[-2,2]$. \qed
\end{example}

Logically, variables in {\fp c} and in {\fp dummyVars} are existentially
quantified variables, inside the RIS. Operationally, they are treated as dummy
variables, i.e., a new instance of the variables in {\fp c} and in {\fp
dummyVars} is created for each application of {\fp F} and {\fp p}. The use of
variables in {\fp dummyVars} will be discussed in Section \ref{locals}.

Given that {\fp Ris} extends {\fp LSet}, {\fp Ris} objects can be used as
logical sets, and all methods of {\fp LSet} are inherited by {\fp Ris}. Some of
these methods, however, are suitably adapted to work with RIS. For example, the
utility method {\fp isBound()} returns true iff the \emph{domain} of the {\fp
Ris} object is bound to some value.

RIS can be expanded into the corresponding extensional sets under certain
conditions.

\begin{definition}[Expandable RIS]
The RIS $\ris{\ct}{D}{\flt}{\ptt}$ is \emph{expandable} if and only if either
$D$ is empty or $D$ contains at least a ground element and the filter $\flt$
does not contain free variables.\qed
\end{definition}

If {\fp R} is an expandable RIS, then the method {\fp R.expand()} returns the
{\fp LSet} object containing the result of the application of the pattern to
each element of the domain that is ground and satisfies the filter. In
particular, if the domain is empty the expansion of the RIS is the empty {\fp
LSet}. If {\fp R} is not expandable, {\fp R.expand()} raises an exception.

\begin{example}\label{ex:expand}
If {\fp R} is the {\fp Ris} object created in Example \ref{ex:Ris obj}, then
the corresponding extensional set {\fp S} is computed and printed as follows:
\begin{alltt}
   LSet S = R.expand().setName("S");
   S.output();
\end{alltt}
\noindent whose execution yields {\fp \_S = \{0,-2,2\}}. \qed
\end{example}

The following are two more examples of RIS that can be written using JSetL
(note that in these examples the RIS patterns are omitted since they are the
same as the corresponding control terms).

\begin{example}[{\fp Ris} objects]\label{ex:Ris obj2}
\mbox{}

\begin{enumerate}
\item[$i.$]
The set of sets, belonging to $D$, containing a given set $A$ (i.e., $\{S : D |
A \subseteq S\}$):
\begin{alltt}
   LSet A = new LSet("A");
   LSet S = new LSet(), D = new LSet();
   Ris R = new Ris(S,D,A.subset(S));
\end{alltt}
\noindent


\item[$ii.$]
The set of ordered pairs $(S,m)$ belonging to $D$, where $S$ is a set and $m$
is its cardinality, provided $m$ is greater than $1$ (i.e., $\{(S,m) : D | m =
\lvert S \rvert \land m > 1\}$):
\begin{alltt}
   LSet S = new LSet(), D = new LSet();
   IntLVar m = new IntLVar();
   Ris R = new Ris(new LPair(S,m),D,S.size(m).and(m.gt(1)));
\end{alltt}
\noindent
where the intuitive meaning of {\fp S.size(m)} is $m = \lvert S \rvert$. \qed
\end{enumerate}
\end{example}

\subsection{RIS constraints}

In this section we show how the atomic set constraints provided by JSetL are
extended to work with RIS as well. A complete list of all JSetL constraint
methods can be found in the JSetL User's Manual \cite{jsetlman}.

\begin{definition}[Atomic RIS constraints]
An \emph{atomic RIS constraint} is an expression of one of the forms:
\begin{enumerate}
\item[$-$] {\fp o.$op$\,(R)}, $op \in \{{\fp in},
{\fp nin}\}$, where {\fp R} is a {\fp Ris} and {\fp o} any logical object;
\item[$-$] {\fp S1.$op$\,(S2)}, $op \in \{{\fp eq}, {\fp subset},
{\fp disj}, {\fp neq}, {\fp nsubset}, {\fp ndisj} \}$;
\item[$-$] {\fp S1.$op$\,(S2,S3)}, $op \in \{ {\fp union}, {\fp inters}, {\fp diff}, {\fp nunion},
{\fp ninters}, {\fp ndiff} \}$,
\end{enumerate}
\noindent where {\fp S1}, {\fp S2} and {\fp S3} are either {\fp LSet} objects
or objects of the Java class {\fp Set}, and at least one of them is a {\fp Ris}
object. \qed
\end{definition}

The meaning of these methods is the one naturally associated with their names:
{\fp eq} and {\fp neq} stand for equality and inequality; {\fp subset} and {\fp
nsubset}, for set inclusion and not inclusion, and so on.

Atomic constraints can be combined using the methods {\fp and} and {\fp or},
whose intuitive meaning is logical conjunction and disjunction, respectively.

\begin{definition}[JSetL constraints]\label{constraint}
A \emph{JSetL constraint} is either an atomic constraint (in particular, an
atomic RIS constraint), or an expression of one of the forms:
\begin{enumerate}
\item[$-$] {\fp C1.and\:(C2)}
\item[$-$] {\fp C1.or\:(C2)}
\end{enumerate}
\noindent where {\fp C1} and {\fp C2} are (recursively) JSetL constraints. Both
atomic and general JSetL constraints are instances of the class {\fp
Constraint}. \qed
\end{definition}


\begin{example}[RIS constraints] \label{ex:RIS_constraints}
If {\fp R} is the {\fp Ris} object created in Example \ref{ex:Ris obj}, then
the following are possible RIS constraints posted on {\fp R}:
\begin{alltt}
   LSet S = LSet.empty().ins(-2,0,2);
   solver.add(R.eq(S));    // \{x:[-2,2] \(|\) x mod 2 = 0 \(@\) x\} = \{-2,0,2\}
   LVar y = new LVar(1);
   solver.add(y.nin(R));   // 1 nin \{x:[-2,2] \(|\) x mod 2 = 0 \(@\) x\}
\end{alltt}

\noindent The same can be obtained by posting a conjunction of the two atomic
constraints:
\begin{alltt}
   solver.add(R.eq(S).and(y.nin(R)));    \qed
\end{alltt}
\end{example}

\subsection{Constraint solving with RIS}

RIS constraints are solved by the JSetL solver using the same
technique adopted for all other constraints. Basically, RIS
constraints are reduced to a solved form using the rewrite rules
developed for the theory of RIS presented in Section \ref{theory}.
In order to account for RIS, the solved form returned by the solver
is extended accordingly. The following notion is crucial in the
definition of solved form for RIS.

\begin{definition}[Variable-RIS]
A RIS is a \emph{variable-RIS} if its domain is an uninitialized logical set;
otherwise it is a \emph{non-variable-RIS}. \qed
\end{definition}

\begin{definition}\label{solved form} (RIS constraints in solved form)
Let {\fp R}, {\fp R1}, {\fp R2} be variable-RIS, {\fp X} an uninitialized {\fp
LSet} object, {\fp V1}, {\fp V2} either variable-RIS or uninitialized {\fp
LSet} objects, {\fp o} any logical object, and $\emptyset$ the {\fp LSet}
object representing the empty set. An atomic RIS constraint of a JSetL
constraint $C$ is in \emph{solved form} if it has one of the following forms:
\begin{enumerate}
\item[$-$] {\fp X.eq(R)}, and {\fp X} does not occur in the other constraints of $C$
\item[$-$] {\fp R.eq($\emptyset$)} or {\fp $\emptyset$.eq(R)}
\item[$-$] {\fp R1.eq(R2)}
%
%
\item[$-$] {\fp o.nin(R)}
\item[$-$] {\fp R1.disj(V1)} or {\fp V1.disj(R1)}
\item[$-$] {\fp R1.union(V1,V2)} or {\fp V1.union(R1,V2)}
or {\fp V1.union(V2,R1)}.   \qed
\end{enumerate}
\end{definition}

Note that all RIS occurring in a RIS constraint in solved form are
variable-RIS.

Intuitively, the key idea behind the rewriting rules for RIS is a sort of
\emph{lazy partial evaluation} of RIS. That is, a RIS object is treated as a
block until it is necessary to identify one of its elements. When that happens,
the RIS is transformed into an extensional set whose element part is the
identified element and whose set part is the rest of the RIS. At this point,
classic set constraint rewriting (in particular set unification) can be
applied. For example, the RIS $\{x : \{z \plus D\} | \flt @ \ptt\}$ will be
rewritten, in general, to the extensional set $\{\ptt(z) \plus \{x : D | \flt @
\ptt\}\}$, provided $\flt(z)$ holds.

According to Cristi\'a and Rossi \cite{DBLP:conf/cade/CristiaR17}, a $\LRIS$ constraint where all
its atomic constraints are in solved form is satisfiable w.r.t. the
interpretation structure (i.e., there exists an assignment of values to all
variables of the constraint that makes it true in the considered
interpretation). Since the rewriting rules applied by the solver to its input
constraint are proved to preserve the set of solutions of the input formula,
then the ability of the solver to produce a solved form constraint guarantees
the satisfiability of the original constraint. Conversely, if the solver
detects a failure, then the original constraint is unsatisfiable.

It is important to observe that if the input constraint is satisfiable, then
the collection of the generated solved form constraints constitute
a finite representation of all the concrete (or ground) solutions of the given
constraint. In JSetL all the computed constraints in solved form can be
displayed using some utility methods, such as {\fp output} and {\fp showStore}.

The following are two examples of RIS constraints along with the
answer computed by the JSetL constraint solver.
\begin{example}[RIS constraint solving]\label{ex:jsetl_solving}
\mbox{}
\begin{enumerate}
\item[$i.$]
Executing the code (cf. the second formula of Example \ref{ex:solving}):
\begin{alltt}
   IntLVar x = new IntLVar("x");
   LSet D = new LSet("D");
   Ris R = new Ris(x,D,x.mod(2).eq(0));
   LSet S = LSet.empty().ins(2,4,6);   // S = \{2,4,6\}
   solver.solve(S.eq(R));
   D.output();
   solver.showStore();
\end{alltt}
\noindent will produce the output:
\begin{alltt}
   _D = \{2,4,6/_N1\}
   Store: \{_x : _N1 \(|\) _N2 = 0 AND _N2 = _x mod 2 \(@\) _x \} = \{\}
\end{alltt}
 \noindent meaning that the given constraint is satisfiable, with
{\fp D} bound to {\fp \{2,4,6/\_N1\}} and the constraint store containing a
solved form RIS constraint involving {\fp \_N1}, where {\fp \_N1} and {\fp
\_N2} are fresh uninitialized logical objects of the proper type.

This result is obtained through the following rewriting/unification steps. The
posted constraint is (in abstract notation) $\{x : D | x\:\:\Mod\:\:2 = 0
\mathbin{\bullet} x\} = \{2,4,6\}$; this is rewritten by the rule dealing with
equality between a variable-RIS and an extensional set. According to this rule,
this equality is rewritten to $D = \{n \plus N\} \land n\:\:\Mod\:\:2 = 0 \land
n=2 \land \{x : N | x\:\:\Mod\:\:2 = 0 \mathbin{\bullet} x\} = \{4,6\}$, where
$n$ and $N$ are fresh variables, and $D = \{n \plus N\}$ means that $n$ must
belong to $D$, while $\{x : N | x\:\:\Mod\:\:2 = 0 \mathbin{\bullet} x\}$
represents the ``rest'' of the RIS. The same rule is applied repeatedly to the
equality between the new variable-RIS and the rest of the extensional set,
until the extensional set is rewritten to the empty set. In that case, all
atomic constraints in the store are in solved form and the solver can stop,
returning the result shown above.
\item[$ii.$]
Executing the code (cf. the third formula of Example \ref{ex:solving}):
\begin{alltt}
   LSet A = new LSet(), B = new LSet(), C = new LSet();
   IntLVar x = new IntLVar();
   solver.add(A.eq(new Ris(x,new LSet(),x.neq(0))));
   solver.add(A.union(B,C).and(A.disj(C)));
   solver.add(A.neq(LSet.empty()));
   solver.solve();
\end{alltt}
 \noindent causes the solver to detect a failure, raising
the exception {\fp Failure}. \qed
\end{enumerate}
\end{example}


\section{Declarative Programming with RIS}\label{declarative}

Intensional sets represent a powerful tool for supporting a declarative
programming style, as pointed out for instance by Dovier et al. \cite{Dovier00}. In this (and
the next) section we provide some evidence for this claim by showing a number
of simple programming examples, using JSetL's facilities for RIS creation and
manipulation.

\subsection{Using RIS to define Restricted Universal Quantifiers}

A first interesting application of RIS to support declarative programming is to
represent \emph{Restricted Universal Quantifiers} (RUQ). The RUQ:
 $$\forall x \in D: \flt(x)$$
can be easily implemented by using a RIS as follows:\footnote{This is formally
proved by observing that $\forall x \in D: \flt(x)$ is just a
notation for $\forall x(x \in D \implies \flt(x))$ and $\forall x(x \in
D \implies \flt(x)) \iff \forall x(x \in D \implies x \in D \land \flt(x)) \iff
\forall x(x \in D \implies x \in \{x:D | \flt(x)\}) \iff D \subseteq \{x:D |
\flt(x)\}$.}
 $$D \subseteq \{x:D | \flt(x)\}$$
Intuitively, solving this formula amounts to check whether $\flt(x)$ holds for
all $x$ in $D$.

RUQ are made available in JSetL by exploiting the JSetL constraint {\fp subset}
applied to RIS. The next two examples are Java programs that solve
simple---though not trivial---problems using JSetL with RIS. Basically, their
solution is expressed declaratively as a formula using RUQ.

\begin{example}\label{ex:min}
Compute and print the minimum of a set of integers {\fp S}.
\begin{alltt}
   public static LVar minValue(LSet S) throws Failure \{
      IntLVar x = new IntLVar(), m = new IntLVar();
      Ris R = new Ris(x,S,m.le(x));
      solver.add(m.in(S).and(S.subset(R)));
      solver.solve();
      return m;
   \}
\end{alltt}
\noindent The method {\fp minValue} posts the constraint $m \in S
\land S \subseteq \{x:S | m \leq x\}$. The solver, non-deterministically binds
a value from $S$ to $m$ and then it checks if the property $m \leq x$ is true
for all elements $x$ in $S$. If this is not the case, the solver backtracks and
tries a different choice for $m$. A possible call to this method is:
\begin{alltt}
   Integer[] sampleSetElems = \{8,4,6,2,10,5\};
   LSet A = LSet.empty().insAll(sampleSetElems);
   LVar min = minValue(A).setName("min");
   min.output();
\end{alltt}
\noindent and the printed answer is \verb!_min = 2!. \qed
\end{example}

It is important to observe that operations on logical sets, including RIS, are
dealt with as constraints. This implies, among others, that it is possible to
compute even with partially specified sets \cite{Bergenti2011}. For example, the set
passed to the method {\fp minValue} can be {\fp \{8,z,4,6\}}, where {\fp z} is
an uninitialized logical variable, or it can contain an unknown part, e.g.,
{\fp \{8,4\plus S\}} where {\fp S} is an uninitialized {\fp LSet} object, or
even it can be simply an uninitialized {\fp LSet} object. In all cases the
JSetL solver is able to check the given constraints and possibly to find a
solution for them. For instance, if {\fp A} is the set {\fp \{8,z,4,6\}}, then
the call {\fp minValue(A)} will non-deterministically generate two distinct
answers, one with ${\fp min} = {\fp z}, {\fp z} \leq 4$, and another with ${\fp
min} = 4, {\fp z} \geq 4$.

This observation shows that this way of iterating over all elements of a
set is not the same as using, for instance, a conventional Java iterator. In
fact, using a RIS allows us to declaratively express a property over all
elements of the set even if they are only partially known.

Another example that shows the use of RIS to define a universal quantification
in a declarative way is the following simple instance of the well-known map
coloring problem.

\begin{example} \label{ex:coloring}
Given a set of $n$ regions $Rg$, a cartographic map $Mp$ of regions in $Rg$, 
and a set $Cl$ 
of $m$ colors, $n, m \ge 1$, find an assignment of colors to the regions such
that no two neighboring regions have the same color.
Each region in the set $Rg$ can be represented as a distinct logical variable
and a map as a set of unordered pairs (hence, sets) of variables representing
neighboring regions. An assignment of colors to regions is represented by an
assignment of values (i.e., the colors) to the logical variables representing
the different regions.

\begin{alltt}
   public static void coloring(LSet Rg, LSet Mp, LSet Cl)
   throws Failure \{
      solver.add(Rg.subset(Cl));
      LSet P = new LSet();
      Ris R = new Ris(P, Mp, P.size(2));
      solver.add(Mp.subset(R));
      solver.solve();
   \}
\end{alltt}

\noindent The method {\fp coloring} posts the constraint $Rg \subseteq Cl \land
Mp \subseteq \{P : Mp | \lvert P \rvert = 2 \}$. The first conjunct exploits
the {\fp subset} constraint to non-deterministically assign a value to all
variables in {\fp regions}. The second conjunct requires that all pairs of
regions in the map have cardinality equal to 2, i.e., all pairs have distinct
components. If {\fp coloring} is called, for instance, with {\fp Rg} $=$ {\fp
\{r1,r2,r3\}}, {\fp r1}, {\fp r2}, {\fp r3} uninitialized logical variables,
{\fp Mp} $=$ {\fp \{\{r1,r2\},\{r2,r3\}\}}, and {\fp Cl} $=$ {\fp \{"red",}
{\fp "blue"\}}, the invocation terminates with success, and {\fp r1}, {\fp r2},
{\fp r3} are bound to {\fp "red"}, {\fp "blue"}, {\fp "red"}, respectively
(actually, also the other solution which binds {\fp r1}, {\fp r2}, {\fp r3}  to
{\fp "blue"}, {\fp "red"}, {\fp "blue"}, respectively, can be computed through
backtracking). \qed
\end{example}

The method {\fp coloring} uses a pure ``generate \& test'' approach; hence it
quickly becomes very inefficient as soon as the map becomes more and more
complex. However, it may represent a first ``prototype'' whose implementation
can be subsequently refined, without having to change its usage. For example,
the coloring problem can be, alternatively, modelled in terms of Finite Domain
(FD) constraints, and the method {\fp coloring} can be implemented by
exploiting the more efficient FD solver provided by JSetL \cite{jsetlman}. On the
other hand, as already noted for Example \ref{ex:min}, the general formulation
presented here allows the method {\fp coloring} to be immediately exploitable
also to solve other related problems, such as, for instance, given a map and a
set of unknown colors (actually, uninitialized logical variables), find whether
the colors are enough to obtain an admissible coloring of the map and, if this
is the case, which constraints the colors must obey.

\medskip

Solving an equality such as $\{x : \{d/A\} | \flt @ \ptt\} = \e$ requires to
check that the filter $\flt$ is false for all elements in $\{d/A\}$, i.e.,
$\forall x \in \{d/A\}: \lnot \flt$. This restricted universal quantification
is implemented through recursion, by rewriting $\{x : \{d/A\} | \flt @ \ptt\} =
\e$ to $\lnot \flt(d) \land \{x : A | \flt @ \ptt\} = \e$.

The next program illustrates the use of RIS to exploit this kind of universal
quantification.

\begin{example}\label{ex:prime}
Check whether {\fp n} is a prime number or not.

\begin{alltt}
   public static Boolean isPrime(int n) \{
      if (n <= 1) return false;
      IntLVar x = new IntLVar();
      Ris R = new Ris(x,new IntLSet(2,n/2),new IntLVar(n).mod(x).eq(0));
      return solver.check(R.eq(LSet.empty()));
   \}
\end{alltt}

\noindent The method {\fp isPrime} posts the constraint
 $\{x : [2,n/2] | n\: \Mod \: x=0\} = \e$.
The equality between the RIS and the empty set ensures that there is no $x$ in
the interval $[2,n/2]$ such that $n\: \Mod \: x = 0$ holds. If, for instance,
{\fp n} is $101$, then the call to {\fp isPrime} returns {\fp true}. The
method {\fp check()} is used to check constraint satisfiability; {\fp
s.check()} differs from {\fp s.solve()} in that the latter raises an exception
if the constraint in the constraint store of {\fp s} is unsatisfiable, whereas
the former returns a Boolean value indicating whether the constraint is
satisfiable or not. \qed
\end{example}

\subsection{Using RIS to define partial functions}\label{pfun}

Another notable application of RIS is using them to represent \emph{(partial)
functions} as sets. In general, a RIS of the form $\{x : D | F @ (x,f(x))\}$,
where $f$ is any function definable in the underlying language, represents a
partial function with domain $D$. In fact, such a RIS denotes a set of ordered
pairs as its pattern is an ordered pair; besides, it is a (partial) function
because each of its first components never appears twice, since they belong to
the set $D$.

Given that RIS are sets, and partial functions can be represented as RIS, then
partial functions can be evaluated, compared and point-wise composed through
standard set operators; moreover, the inverse of a function can also be
computed by means of constraint solving. The following examples illustrate
these ideas in the context of Java, using JSetL.

\begin{example}\label{ex:square}
The square function of an integer $n$.
\begin{alltt}
   IntLVar x = new IntLVar();
   LSet D = new LSet();
   Ris sqr = new Ris(x,D,Constraint.truec(),new LPair(x,x.mul(x)));
\end{alltt}

\noindent where {\fp Constraint.truec()} is a static method of the class {\fp
Constraint} returning an always true constraint. {\fp sqr} defines the set of
all ordered pairs $(x,x*x)$, with $x$ belonging to a set {\fp D}. This function
can be ``evaluated'' in a point $n$, and the result sent to the standard
output, by executing the following code:
\begin{alltt}
   IntLVar y = new IntLVar("y");
   solver.solve(new LPair(n,y).in(sqr));
   y.output();
\end{alltt}
\noindent that is, {\fp y} is the image of {\fp n} through function {\fp sqr}.
If, for instance, {\fp n} has value 5, then the printed result is {\fp \_y =
25}. Note that the RIS domain, {\fp D}, is left underspecified as a variable.
\qed
\end{example}

Hence, RIS provides, in a sense, a facility to go from the static
definition of a Java function to a set theoretic view of this function. More
precisely, if {\fp T$_1$ f(T$_2$ x)} is a Java method,
where {\fp T$_1$} and {\fp T$_2$} are Java types, then its corresponding set
theoretic counterpart is given by the RIS $\{x : D | F @ (x,f(x))\}$, where
$\{x | x \in D \land F\} \subseteq T_2$ is the \emph{domain of definition} of
$f$.

A first advantage of this approach is that, as usual in declarative
programming, there is no real distinction between inputs and outputs.
Therefore, the same RIS of Example \ref{ex:square} can be used also to
calculate the inverse of the square function, that is the square root of a
given number. To obtain this, it is enough to replace the call to {\fp solve} in Example
\ref{ex:square} with the following new call:
\begin{alltt}
   solver.solve(new LPair(y,n).in(sqr));
\end{alltt}

\noindent If, for instance, {\fp n} has value 25, then the computed result is
{\fp \_y = unknown -- Domain: \{-5, 5\}}, stating that the possible values for
{\fp \_y} are {\fp -5} and {\fp 5}.

\medskip

An interesting aspect of using RIS for defining functions is
that RIS are sets and sets are data. Thus, we have a simple way to deal with
functions as data. In particular, since {\fp Ris} objects can be passed as
arguments to a function, we can use RIS to write generic functions that take
other functions as their arguments. The following is an example illustrating
this technique.

\begin{example}\label{ex:mapList}
The following method takes as its arguments an array of integers {\fp A} and a
function {\fp f} and updates {\fp A} by applying {\fp f} to all its elements:
\begin{alltt}
   public static void mapList(Integer[] A,LSet f) throws Failure \{
      for(int i=0; i<A.length; i++) \{
         IntLVar y = new IntLVar();
         solver.solve(new LPair(A[i],y).in(f));
         A[i] = y.getValue();
      \}
   \}
\end{alltt}

\noindent If, for instance, {\fp elems} is the array with values {\fp
\{3,5,7\}} and {\fp f} is the {\fp Ris} object {\fp sqr} of Example
\ref{ex:square}, then the call {\fp mapList(elems,sqr)} will yield the
modified array {\fp \{9,25,49\}}. \qed
\end{example}

The use of RIS to represent partial functions allows them to be considered as
Java first class citizens, as usual in functional languages. As such, the use
of RIS constitutes a viable alternative to other functional facilities recently
introduced in Java, such as lambda expressions. For example, the method {\fp
mapList} could be written in pure Java as (see
\ref{app:java_programs} for the complete Java code):

\begin{alltt}
    public static void mapList(Integer[] A,FunctionInt f)  \{
        for(int i=0; i<A.length; i++) \{
            A[i] = f.apply(A[i]);
        \}
    \}
\end{alltt}
\noindent where {\fp FunctionInt} is a Java {\fp interface} containing the
only method {\fp apply}, from {\fp int} to {\fp int}, and its invocation using
lambda expression can be {\fp mapList(elems,x -> x*x)}. Thus the difference
between the two solutions is mainly methodological: using a set-theoretical
approach in one case; using a functional approach in the other.

A positive aspect of the set-theoretical approach, however, is its
declarativity, in particular the possibility to not distinguish between inputs
and outputs. Thus, for instance, replacing the last two
statements of {\fp mapList} with the single statement:
\begin{alltt}
   if (solver.check(new LPair(y,A[i]).in(f)))
      System.out.println(A[i]);
\end{alltt}
\noindent (i.e., swapping the two parameters of the {\fp LPair} object in
the {\fp in} constraint) provides an easy solution to the problem of printing
all numbers in the given array {\fp A} that are the squares of some integer
number.

\subsection{Using JSetL as a theorem prover}\label{theo}

As already observed, in JSetL operations on logical objects are dealt with as
constraints. Thus it is possible to compute with logical objects, such as {\fp
LSet} and {\fp Ris} objects, even if they are only partially specified or
completely unknown. In particular, we can use the JSetL solver to check
satisfiability of very general formulas involving both extensional and
intensional sets, in a similar way to what is done with theorem provers.

\begin{example} \label{ex:inters}
Check the property $C = A \cap B \iff C = \{x : A | x \in B\}$. This is proved
in JSetL by showing that the formula $\Cap(A,B,C) \land R= \{x : A | x \in B @
x\} \land R \neq C$ is false.
\begin{alltt}
   LSet A = new LSet(), B = new LSet(), C = new LSet();
   solver.add(A.inters(B,C));      // the constraint C \(=\) A \(\cap\) B
   LVar x = new LVar();
   Ris R = new Ris(x,A,x.in(B));   // R \(=\) \{x:A \(|\) x \(\in\) B\}
   solver.add(R.neq(C));           // the constraint R \(\neq\) C
\end{alltt}

\noindent Calling {\fp solver.solve()} causes the exception {\fp Failure} to be
thrown (i.e., the formula is found to be $\false$).  \qed
\end{example}

The next example shows that JSetL can be used as a prover for a non-trivial
fragment of first-order logic with quantifiers.

\begin{example} \label{ex:all_positive}
The formula $(\forall x \in S : x > 0) \land -1 \in S$ can be written in JSetL
with RIS as follows:
\begin{alltt}
   LSet S = new LSet();
   IntLVar x = new IntLVar();
   Ris R = new Ris(x,S,x.gt(0));
   Constraint C = S.subset(R).and(new IntLVar(-1).in(S));
\end{alltt}
\noindent and can be proved to be unsatisfiable by posting and solving the
constraint {\fp C} by using the JSetL solver:
\begin{alltt}
   solver.solve(C);   \normalsize{\qed}
\end{alltt}
\end{example}


\section{Extended RIS}\label{extensions}

To guarantee that the constraint solver is indeed a decision procedure a number
of restrictions are imposed on the form of RIS
\cite{DBLP:conf/cade/CristiaR17}. Specifically: $(i)$ the control term and
pattern of RIS are restricted to be of specific forms---see Definition
\ref{def:admissible}; $(ii)$ the filter of RIS cannot contain ``local''
variables, i.e., existentially quantified variables declared inside the RIS,
besides those in the control term; and $(iii)$ recursively defined RIS such as
$X = \{x : D | \flt(X) @ \ptt\}$ are not allowed.\footnote{Note that, on the
contrary, a formula such as $X = \{D(X) | \flt @ \ptt\}$ is an admissible
constraint, and it is suitably handled by the JSetL solver.}

Although compliance with these restrictions is important from a theoretical
point of view, in practice there are many cases in which they can be
(partially) relaxed without compromising the correct behavior of programs using
RIS.

In this section we show how JSetL extends the language of RIS presented in
Section \ref{theory}, by relaxing all the above mentioned restrictions. We
also show, through a number of simple examples, that the availability of these
new features can considerably enhance the expressive power of the language.

\subsection{RIS with general patterns}

As noted by Cristi\'a and Rossi \cite{DBLP:conf/cade/CristiaR17}, a condition for patterns to
guarantee correctness and completeness of the constraint solving procedure is
for patterns to be bijective functions. All the admissible patterns of $\LRIS$
are bijective patterns. Besides these, however, other terms can be bijective
patterns. For example, $x+n$, $n$ constant, is also a bijective pattern, though
it is not allowed in $\LRIS$. Conversely, $x*x$ is not bijective as $x$ and
$-x$ have $x*x$ as image, although $(x,x*x)$ is indeed a bijective pattern
allowed in $\LRIS$.

Unfortunately, the property for a term to be a bijective pattern cannot be
easily syntactically assessed. Thus Cristi\'a and Rossi \cite{DBLP:conf/cade/CristiaR17} adopt a more
restrictive definition of admissible pattern. However, from a
practical point of view, as in JSetL, we can admit also more general patterns.
If the expression used in the RIS pattern defines a bijective function, then
dealing with the RIS is safe (i.e., the answer computed by the solver is fully
reliable); otherwise, it is not safe in general.

Specifically, RIS patterns in JSetL can be any logical object. In particular,
we allow patterns to be integer logical expressions involving variables
occurring in the RIS control term. Users are responsible of using bijective
patterns.

Actually, bijectivity of the function defining the pattern must be assessed not
in general, but with respect to the RIS domain restricted to those values
satisfying the RIS filter. The following is an example using a RIS whose
pattern is a bijective function in its domain of interest.

\begin{example}\label{ex:bijective}
Compute the set of squares of all even numbers in $[1,10]$.
\begin{alltt}
   IntLVar x = new IntLVar();
   Ris R = new Ris(x,new IntLSet(1,10),x.mod(2).eq(0),x.mul(x));
   \hfill //R = \{x:[1,10] \(|\) x mod 2=0 \(@\) x*x\}
   LSet Sqrs = R.expand();
\end{alltt}

\noindent Executing this code will bind {\fp Sqrs} to {\fp \{4,16,36,64,100\}}.
\qed
\end{example}

Conversely, if the domain of the RIS in Example \ref{ex:bijective} is the
interval $[-10,10]$, then the pattern is not bijective and the
computed answer will be, in general, not safe. For
instance, the simple formula  $\{x : D | \true @ x*x\} = \{4\} \land 2 \in D
\land -2 \in D$ is found to be unsatisfiable, even if it has the trivial
solution $D = \{-2,2\}$. Intuitively, the problem originates from the fact that
when processing $\{x : D | \true @ x*x\} = \{4\}$, where $D$ is a variable, the
element $4$ is extracted from $\{4\}$ and the value $x$ for which $4 = x*x$ is
added to $D$; if more than one $x$ has $4$ as its image, then only one of them
is added to $D$; hence, either $2 \in D$ or $-2 \in D$ fails.

\subsection{RIS with dummy variables}\label{locals}

Allowing existentially quantified variables in RIS raises major problems when
the formula representing the RIS filter has to be negated during RIS constraint
solving (basically, negation of the RIS filter is necessary to assure that any
element that does not satisfy the filter does not belong to the RIS itself). In
fact, this would require that the solver is able to deal with possibly complex
universally quantified formulas, which is usually not the case (surely, it is
not the case for the JSetL solver).\footnote{Variables occurring in the control
term are also (implicitly) existentially quantified. However, since they are
required to take their values from the RIS domain, negation of the RIS filter
for such variables turns out to be a form of restricted universal
quantification which is conveniently implemented through recursion, by
extracting one element at a time from the RIS domain.} Thus, to avoid such
problems \emph{a priori}, in $\LRIS$ the RIS filter cannot contain any explicit
existentially quantified variable.

However, as already observed for RIS patterns, in practice there are cases in
which we can relax restrictions on RIS without losing the ability to correctly
deal with more general RIS constraints.

Thus, in JSetL, we allow the user to specify that some (logical) variables in
the RIS filter are indeed local variables. This is achieved by using the fifth
argument of the {\fp Ris} constructor, which accepts a sequence of logical
objects that the user wants to be treated as existentially quantified (or
dummy) variables.

Since problems with existentially quantified variables inside RIS are generated
by the possible use of negation in RIS filters, and since not all rewrite rules
dealing with RIS constraints require such negation to be applied, then there
are cases in which dummy variables in RIS can be used safely. For this reason, the solver raises an exception {\fp UnsafeRisException} every
time a rewrite rule requiring the negation of the RIS filter is applied to a
RIS with dummy variables.
In this way, if the exception is not risen the computed answer is
guaranteed to be correct. In the following example, the solver is able to compute a correct answer
although the RIS declares a dummy variable.

\begin{example}\label{ex:parameter}
If $S$ is a set of ordered pairs and $D$ is a set, then the subset of $S$ where
all the first components belong to $D$ (i.e., the \emph{domain
restriction} of relation $S$ to $D$) can be defined as $\{x : D | \exists y
((x,y) \in S @ (x,y))\}$, where $D$ and $S$ are free variables, while $x$ and
$y$ are existentially quantified variables. This is implemented in JSetL by the
following declarations:

\begin{alltt}
   LSet S = new LSet("S"), D = new LSet("D");
   LVar x = new LVar(), y = new LVar();
   Ris R = new Ris(x,D,new LPair(x,y).in(S),new LPair(x,y),y);
\end{alltt}

\noindent If we execute:
\begin{alltt}
   solver.solve(new LPair(1,2).in(R).and(new LPair(3,4).in(R)));
   D.output(); S.output();
\end{alltt}
\noindent 
then the program terminates with success, printing:
\begin{verbatim}
   _D = {1,3/_N1}
   _S = {(1,2),(3,4)/_N2}
\end{verbatim}
\noindent where {\fp \_N1} and {\fp \_N2} are fresh uninitialized logical sets.
\qed
\end{example}

In the above example, {\fp y} is a dummy variable. If {\fp y} is not declared
as dummy, then the same call to {\fp solver.solve} will terminate with failure,
since {\fp y} is dealt with as a free variable and the first constraint, $(1,2)
\in {\fp R}$, binds {\fp y} to $2$ so that the second constraint $(3,4) \in
{\fp R}$ fails.

It is worth noting that many uses of dummy variables can be avoided by a proper
use of the control term and pattern of a RIS. For example, the RIS of Example
\ref{ex:parameter} can be replaced by the RIS without dummy variables $\{(x,y)
: S | x \in D @ (x,y)\}$. Hence,
allowing control terms and patterns for RIS to be any logical object can also
be useful to alleviate the problem of existentially quantified variables in RIS
filters.

\subsection{Recursive RIS}

The class {\fp Ris} extends the class {\fp LSet}. Hence it is possible to use
{\fp Ris} objects inside the RIS filter formula in place of {\fp LSet} objects.
This allows, among other things, to define \emph{recursive restricted
intensional sets} (RRIS).

The presence of recursive definitions may compromise the finiteness of RIS and
hence the decidability of RIS formulas.
Therefore RRIS are prohibited in the base language of RIS, $\LRIS$. In
practice, however, their availability can considerably enhance the expressive
power of the language and hence RRIS are allowed in the extended version of
$\LRIS$ implemented in JSetL.

Using RRIS in JSetL may cause the constraint solving procedure to run forever.
On the other hand, correctness is preserved: if the solver terminates the
computed answer is guaranteed to be correct. Ensuring termination is the
responsibility of the programmer.

As shown in Section \ref{pfun} a function $f$ can be defined as a set of
ordered pairs $G_f = \{x:D | \flt @ (x,f(x))\}$, for some filter $\flt$ and
domain $D$. A call to $f$, e.g., $y=f(x)$, is simply expressed as a set
membership predicate over the set defining $f$, i.e., $(x,y) \in G_f$. A call
to $f$ in the filter of the RIS defining $f$ itself
is a recursive call to $f$. For example, the well known factorial function
${\fp fact}(x)$ can be defined as a recursive RIS as follows:
$${\fp fact} = \{(0,1) \plus \{x:D | \exists z(x > 0 \land (x-1,z) \in {\fp fact}
@ (x,z*x))\}.$$

\noindent Note that the domain of the RIS is left underspecified, and recursion
is simply expressed as $(x-1,z) \in {\fp fact}$, meaning that $z$ is the
factorial of $x-1$. Also note that the base case of the recursive definition of
{\fp fact} is simply added as a known element, $(0,1)$, to the set being
defined.

Such kind of recursive definitions are directly supported by the implementation
of RIS in JSetL.

\begin{example}[Factorial of a number $x$]\label{ex:fact}
\mbox{}
\begin{alltt}
   LSet fact = new LSet();
   IntLVar x = new IntLVar(), z = new IntLVar();
   Constraint C = x.gt(0).and(new LPair(x.sub(1),z).in(fact));
   Ris R_fact = new Ris(x,new LSet(),C,new LPair(x,z.mul(x)),z);
   solver.add(fact.eq(R_fact.ins(new LPair(0,1))));
\end{alltt}
\noindent where {\fp z} is a dummy variable which is used to contain ${\fp
fact}(x - 1)$. If we conjoin, for example, the constraint {\fp new
LPair(5,ff).in(fact)}, where {\fp ff} is an uninitialized {\fp IntLVar}, and
ask the solver to solve the current constraint, then the solver will return
{\fp ff = 120} (while {\fp D} will be bound to {\fp \{5,4,3,2,1/\_N\}}).
Conversely, if we conjoin the constraint {\fp new LPair(n,120).in(fact)}, where
{\fp n} is an uninitialized {\fp IntLVar}, then the solver will return {\fp n =
5} (and {\fp D} is bound again to {\fp \{5,4,3,2,1/\_N\}}). \qed
\end{example}

The following is another non-trivial example using both recursive RIS and dummy
variables.

\begin{example}[Reachable nodes]\label{ex:graph}
Given a directed acyclic graph $G = \langle N,E \rangle$, where $N$ is a
not empty set of nodes and $E$ the set of directed edges over $N$, and a
node $n \in N$, compute the set $R$ of all nodes reachable from $n$ (including
$n$ itself) in an arbitrary long number of steps. Using RIS it is possible to
compute $R$ as follows:
\begin{verbatim}
   public static LSet reachable(LSet N, LRel E, LVar n){
      LVar x = new LVar("x");
      LVar y = new LVar("y");
      LSet R = new LSet();
      Ris R_R = new Ris(x,N,x.eq(n)
                    .or(y.in(R).and(new LPair(y,x).in(E))),x,y);
      R.eq(R_R).check();
      return R_R.expand();
}
\end{verbatim}
 \noindent where {\fp LRel} is a JSetL class, extending {\fp LSet}, that provides
the data abstraction of \emph{binary relation} (i.e., sets of ordered pairs).
The {\fp Ris} object {\fp R\_R} represents the RIS $\{x : N | x = n \lor
\exists y(y \in R \land (y,x) \in E))\}$. The statement {\fp R.eq(R\_R).check()
forces ${\fp R} = {\fp R\_R}$ to hold,} thus making {\fp R\_R} a recursive RIS.
Finally, the last line of {\fp reachable} returns an extensional set containing
the set of nodes reachable from {\fp n}. The code below shows an example of the
usage of the method {\fp reachable}.
\begin{verbatim}
   Integer[] nodesArray = {1,2,3,4,5,6};
   LPair[] edgesArray = {new LPair(1,2), new LPair(1,3),
                         new LPair(2,5), new LPair(4,6)};
   LSet Nodes = LSet.empty().insAll(nodesArray);
   LRel Edges = LRel.empty().insAll(edgesArray);
   LVar start = new LVar(1);
   reachable(Nodes,Edges,start).setName("Reachable").output();
\end{verbatim}
 \noindent
The starting node is {\fp 1} (represented by the logical variable {\fp start}).
The computed output is:
\begin{alltt}
   _Reachable = \{1,2,3,5\}    \normalsize{\qed}
\end{alltt}
\end{example}

RRIS are general enough to allow different types of recursion to be
implemented. Examples of nested recursion (namely, the Ackermann function) and
mutual recursion are shown in \ref{app:jsetl_programs}.


\section{Design and implementation issues}\label{implementation}

As a major design choice we opted to implement declarative programming (in the
form of constraint programming) and RIS in Java by means of a library, and not
as an extension to the language. Hence, all facilities to support declarative
constraint programming are implemented on top of Java by exploiting the
language abstraction mechanisms. In particular, all JSetL logical objects,
including RIS, are implemented as instances of Java classes. The root class of
the JSetL class hierarchy is {\fp LObject}. This class provides general utility
methods that are common to every logical object, such as methods to check
whether an object is initialized or not, to set its external name, and so on.
{\fp LObject} has two subclasses, {\fp LVar} and {\fp LCollection}, where the
latter is the superclass of {\fp LSet} and {\fp LList}.


A major problem with the library-based approach is implementing nondeterminism
and the related backtracking mechanism which are part of the JSetL constraint
solver. Actually, in general, this would be done much better \emph{inside} the
language, via primitive constructs, rather than \emph{on top of} the language.
Our solution is to completely embed non-determinism within the constraint
solver, providing a few library methods, such as {\fp nextSolution}, {\fp
setOf}, {\fp addCoicePoint}, to support interaction between conventional
constructs and the backtracking mechanism used to implement nondeterminism
\cite{DBLP:journals/fuin/RossiB15}.

As concerns RIS, they are easily integrated in the library as a user defined
data abstraction. Their generality, however, allows them to be used also to
express more general abstractions. In particular, RIS can be used to implement
iteration over all elements of a set, as well as to express functions and to
deal with them as data, and to implement various forms of recursion.

A number of techniques are used to overcome possible
performance bottlenecks when dealing with logical sets and RIS. In particular,
when all elements of the involved sets are constants then special ad-hoc rules
are used to process them, instead of using the more general rewriting rules
provided by the general constraint solving procedure.  As concerns RIS, it is
worth mentioning the implementation of a cache mechanism for improving
execution of RIS expansion, as performed by the method {\fp expand} (see Sect.
\ref{ssec:ris}). This method constructs an extensional set from a RIS. It is
invoked either explicitly by the user or implicitly as an internal optimization
of the JSetL solver. Its execution requires to solve constraints and perform
syntactic checks on the {\fp Ris} it is invoked on. Since it may be called
several times on the same (ground) elements of the domain of a RIS, the results
computed for each element are stored into a cache. In order to ensure that this
optimization does not hinder the correctness of the result, the solver takes
care to backtrack the whole cache along with the state of the logical
variables. For example, when applied to the problem of finding the reachable
nodes in a graph from a starting node (see Example \ref{ex:graph}) the cache
mechanism allows us to avoid recomputing the set of reachable nodes from the
same node many times thus considerably reducing the computational
time.\footnote{The algorithm used to decide which cache entry to remove from
the cache when it is full is a variation of the clock algorithm for page
replacement.}

A fundamental Java feature we have exploited in the design and implementation
of JSetL is the possibility of using the generic class {\fp Object} to define
classes containing objects of any type. In particular, the collection of values
contained in an {\fp LSet} are instances of {\fp Object}; thus, elements of a
logical set can be of any type, including {\fp LVar} and {\fp LSet} (for nested
sets); more importantly, elements of a logical set are not required to be all
of the same type so that the same set can contain some (possibly uninitialized)
logical variables, along with constant values, of any type.

Another Java feature which turned out to be very useful in practice is the
\emph{varargs} construct, which allows an arbitrary number of values to be
passed to a method. This feature has been exploited, for instance, in the
method {\fp ins} to allow a logical set of $n$ elements to be constructed by
simply listing all its elements as parameters of the method; and in the {\fp
Ris} constructor to list all dummy variables possibly occurring in the denoted
RIS.

On the negative side, the lack of the operator overloading mechanism in Java
forces us to always adopt the prefixed notation for all new methods, resulting
sometimes in a somewhat cumbersome syntactic notation.

Facilities provided by JSetL coexist with conventional and object-oriented
programming constructs of Java. Actually, the programmer can exploit the
abstraction and program structuring mechanisms of Java to separate declarative
code from the more conventional one. A Java program using JSetL is likely to be
composed of classes whose methods are defined in a declarative way, along with
classes whose methods contain (only) conventional imperative code. Both kinds
of methods can be used and understood as usual Java methods. The declarative
ones, however, also admit a more abstract, declarative reading, which is
essentially based on the declarative computational model of the \CLPSET\
language extended with RIS.



\section{Set-oriented Programming in Practice}\label{practice}

The main goal of JSetL is to allow more readable and reliable programs to be
obtained through the use of set theory. Moreover, as a side effect of its
declarative and CLP nature, JSetL supports some verification techniques as shown in
Section \ref{declarative}.

The basic version of JSetL has been shown to be effectively usable in practice
through a number of simple---though often not trivial---programming examples
that are available on-line at the JSetL website \cite{jsetl}.

In order to provide some evidence to this claim also for the new version
extended with RIS, we show here the results of an empirical assessment for most
of the sample programs shown in the paper. Table \ref{t:overview} shows for
each program the time (in milliseconds) it takes to complete with the given
number of elements in input. A zip archive with all these programs, along with
all the statistical data, is available at the JSetL website.

\begin{table}
\caption{\label{t:overview}Summary of the empirical evaluation (times are in
msec)}
\begin{tabularx}{\textwidth}{Xrrrrrrr}
\hline\hline \multicolumn{1}{c}{\textsc{Example /\
Cardinality}} & \multicolumn{1}{c}{\textsc{0}}    &
\multicolumn{1}{c}{\textsc{1}} & \multicolumn{1}{c}{\textsc{5}}    &
\multicolumn{1}{c}{\textsc{10}} & \multicolumn{1}{c}{\textsc{25}}  &
\multicolumn{1}{c}{\textsc{50}}     & \multicolumn{1}{c}{\textsc{100}}
\\\hline
\textsc{8 Ris creation} &  0.65 &   1.04 & 0.65  & 0.67 & 0.70  & 0.71  & 0.82 \\

\textsc{9 Ris expansion} &  0.32 &  0.61 &   1.03 &   1.51 &  3.38 &  5.61 &   10.99  \\
\textsc{11 Ris constraints} &  0.67 &  0.82 &    0.72  & 0.73   & 1.02 &  0.95 & 1.19 \\
\textsc{12 Ris constraint solving} &  0.77  & 0.95    & 1.76    & 2.86  & 5.30   & 8.79  & 17.26 \\
\textsc{13 Minimum of a set} & 0.26   & 0.40  & 0.53  & 1.22  &1.49   & 4.54  & 7.35 \\
\textsc{15 Prime test} &  0.00 &  0.01 &  0.24 &  0.89 &
6.73 &  28.39 & 104.39
 \\
\textsc{20 Squares of evens}  &  0.22   & 0.78 &   1.26  & 1,79  & 3.77 &  6.83  & 12.32 \\
\textsc{21 Domain restriction}   &  0.11 & 0.20  & 0.58    & 1.52 & 11.74  & 75.70   & 525 \\
\textsc{23 Reachable nodes}  &  - & 0.33 &   1.00 & 4.82 & 88.06 & 815 &  11136 \\

\hline\hline \multicolumn{1}{c}{\textsc{Example /\
Cardinality}} & \multicolumn{1}{c}{\textsc{0}} &
\multicolumn{1}{c}{\textsc{1}}    & \multicolumn{1}{c}{\textsc{5}} &
\multicolumn{1}{c}{\textsc{10}}  & \multicolumn{1}{c}{\textsc{11}}     &
\multicolumn{1}{c}{\textsc{12}}   & \multicolumn{1}{c}{\textsc{13}}
\\\hline
\textsc{14 Map coloring}  & 0.05 &    0.47 &  5.48 &  713.55 &    2036 &  5940 & 17680 \\
%
\textsc{22 Factorial}   & 0.52   & 1.28    & 6.46    & 18.53   & 22.94   & 25.21 & - \\

\hline\hline \multicolumn{1}{c}{\textsc{Example /\
Cardinality}} & \multicolumn{1}{c}{\textsc{0}}  &
\multicolumn{1}{c}{\textsc{1}} & \multicolumn{1}{c}{\textsc{2}}  &
\multicolumn{1}{c}{\textsc{3}} & \multicolumn{1}{c}{\textsc{4}}  &
\multicolumn{1}{c}{\textsc{5}} & \multicolumn{1}{c}{\textsc{6}}
\\\hline
\textsc{7 Permutations} &  0.11 &   0.09 &   0.22 & 1.77
&  28.29 & 625 & 17153
  \\\hline\hline
\end{tabularx}
\end{table}

The tests have been conducted on a laptop with Windows 10, 8Gb of RAM memory,
an SSD and an i7-7700HQ clocked at 2.69GHz. All tests were repeated 20 times,
computing the average execution time for each of them. For some examples a
pseudorandom number generator was used to create the input.

%
%
%
For many of the considered problems the JSetL implementation shows reasonable
execution times on data involving sets up to 100 elements. Unfortunately, for
some problems execution times grow very quickly, actually making it impossible
to test these programs for all the considered cardinalities.\footnote{Note
that: $(i)$ Example \textsc{22 Factorial} did not manage to complete for
cardinalities of 13 and above because of limited arithmetic rather than time;
$(ii)$ the input for \textsc{14 Map coloring} was a (pseudo-)randomly generated
bipartite graph and the solver was asked to provide a 2-coloring of it but
there is no optimization in JSetL that accounts specifically for bipartite
graphs and, in fact, the results shows an exponential computational time
growth.}


However, a comparison with more conventional implementations, in particular
using Java, cannot be done simply by comparing the execution times obtained in
the two scenarios. As a matter of fact, JSetL implementations use rather
unconventional algorithms to solve the given problems, favoring readability and
flexibility (e.g., input-output indistinguishability), at the expense of
efficiency. For example, the JSetL set-theoretic implementation of the method
for solving the permutation problem (see Example \ref{ex:permutations}) uses
set unification for non-deterministically generating all possible assignments
of values to the collection of variables representing a single
permutation.\footnote{Set unification, which is a fundamental modeling tool in
JSetL and which underlies many other set operations, has been shown to be an
NP-hard problem in the general case \cite{DBLP:journals/jlp/DovierOPR96}.}
Certainly, the computational complexity of this algorithm is very bad, and the
execution times for this program may be many orders of magnitude larger than
those of a conventional Java implementation for not trivial input data.

On the other hand, the JSetL implementation is simpler and, possibly, much
closer to the set-theoretic formal definition of the problem and, thus, more
evidently correct. To better realize this, a sample pure Java implementation
for the permutation problem is shown in \ref{app:java_programs}. Moreover, as
already noted, the JSetL implementation usually allows more flexible and
general usages. For example, the program for determining the minimum of a set
of integers (cf. Example \ref{ex:min}) can be used not only to compute the
actual value of the minimum but also, for instance, to compute a finite
representation of all sets having a given minimum, as well as for checking
general properties of minimums. As an example, a simple Java+JSetL program for
proving that the minimum of a set (of integers) is always less or equal than
its maximum is shown in \ref{app:jsetl_programs}.

Thanks to its powerful set abstractions (in particular RIS) JSetL can serve as
an effective bridge between set-based formal specification languages, e.g., Z
or B, and conventional O-O programming, e.g., Java. The following general
methodology can be used to develop concrete Java programs from a set-theoretic
formal specifications, through JSetL.
\begin{enumerate}
\item[$(i)$]
The starting point is a set-theoretic specification written in a formal
specification language, such as Z. Note that the Z specification can be
verified with tools such as \setlog
\cite{DBLP:journals/jlp/DovierOPR96,DBLP:journals/jar/CristiaR20,DBLP:conf/cade/CristiaR17,setlog}
and ProB \cite{Leuschel00}.
\item[$(ii)$]
This specification is used to guide the development of a Java+JSetL program
which is as close as possible to the given specification. In particular most
set-theoretic constructs in the specification language have a direct
translation to some JSetL facility. For example, Z variables are mapped to {\fp
LVar} objects, Z types are mapped to {\fp LSet} objects, Z set comprehension
expressions are mapped to {\fp Ris} objects, and so on. The new classes
recently added to JSetL to support the notions of binary relations and partial
functions, together with the RIS abstraction presented in this paper, are
crucial for faithfully representing set-based specifications.

The program produced in this way can be seen as a first executable prototype,
directly derived from formal specifications (in this sense, one can consider
programs based on JSetL as executable set-based specifications).

At present, translation from the formal specification to the Java+JSetL program
is done by hand, in an informal way. However, it could be easily automated and
the implementation of an automatic tool capable of doing this translation is
currently under development.

Besides, refined prototypes could be obtained from the initial one through
subsequent formal refinements. In fact, for all parts of the Java+JSetL program
that are expressed in pure set-theoretical terms, it would be straightforward
to obtain from them the corresponding logical $\LRIS$ formulas. In this way,
one could use tools like \setlog and ProB to prove that the refinement implies
the more abstract version \cite{hcvs2017}.
%
\item[$(iii)$] The purely set-theoretical solutions using JSetL can be subsequently
replaced by more efficient implementations in pure Java (possibly still using
JSetL, but closer to classical imperative solutions). All methods using logical
objects (such as {\fp LSet}) and JSetL constraints can be reimplemented using
more conventional data structures and operations, such as the {\fp Set} objects
of the Java standard library instead of JSetL logical sets.

Although these transformations might not be done in a formal way, the changes
can be confined to those (single) methods that use logical objects and
constraints and whose execution turns out to be too inefficient. The rest of
the program can continue to use these methods in the same way, with no concern
about their actual implementation.

A significant difference compared to similar approaches in the literature is
that in our proposal the same language (namely, Java) is used both for the
prototypes (obtained from the formal specifications), and for the final
implementation.
%
\item[$(iv)$] In addition, the purely set-theoretical methods using JSetL developed at step
$(ii)$ can be used in conjunction with the corresponding, more efficient, pure
Java methods to provide a sort of validation of the results computed by the
latter. As an example, consider the problem of computing the minimum of a set
of integers. From the set-theoretic specification of the minimum of a set $S$,
i.e., $m \in S \land \forall x \in S: m \leq x$ (step $(i)$), we immediately
get (step $(ii)$) an executable Java+JSetL program (see Example \ref{ex:min})
whose core part is implemented by the following method:
\begin{alltt}
   public static void minPred(LSet S, IntLVar m) throws Failure \{
      IntLVar x = new IntLVar();
      Ris ris = new Ris(x,S,m.le(x));
      solver.add(m.in(S).and(S.subset(ris)));
      solver.solve();
    \}
\end{alltt}
This method is intended to represent the relation between a set $S$ and its
minimum $m$: if $m$ is the minimum of $S$ then the method terminates normally;
otherwise it raises a {\fp Failure} exception. As shown in Table
\ref{t:overview}, the performances of this set-theoretical implementation are
enough to allow us to use it as part of a first concrete running prototype.
Then, the method {\fp minValue} can be replaced (step $(iii)$) by a new more
efficient pure Java method. For example:
\begin{alltt}
   public static Integer minValue(Integer[] A) throws Failure \{
      int min = A[0];
      for(int i=1; i<A.length; i++)
         if (A[i] < min) min = A[i];
      minPred(LSet.empty().insAll(A),new IntLVar(min));
      return min;
   \}
\end{alltt}
The call to {\fp minPred} before the {\fp return} statement is used to validate
the result computed at the end of the {\fp for} loop. If $min$ does not contain
the minimum of $A$ then the {\fp Failure} exception is thrown.

In this sense, the JSetL methods can be used as contracts
\cite{DBLP:books/sp/MeyerTouch2009} or as runtime or reference monitors
\cite{DBLP:journals/cuza/BetarteCLR16,DBLP:journals/cleiej/LunaBCSCG18}.
\end{enumerate}


\section{Related work}\label{related}

As concerns the non-RIS fragment of JSetL, the main distinguishing features of
JSetL w.r.t. other libraries for OO languages allowing set variables and set
constrains, such as Choco and Gecode (just to mention two of the best known),
are: sets in JSetL can contain objects of any type, including uninitialized
logical variables, which allows them to be possibly partially specified;
the use of unification, in particular set unification, which allows, in
conjunction with other facilities, multiple uses of procedures where no real
distinction between inputs and outputs is made;
set constraint solving in JSetL does not require any finite set domain to be
associated with set variables, which can range over the whole universe of set
objects; finally, JSetL allows the user to write nondeterministic programs by
exploiting both the nondeterminism embedded in primitive constraint solving
(e.g., in set unification) and the possibility to define new constraints using
the nondeterminism handling facilities provided by the solver itself. On the
other hand, such generality and flexibility provided by JSetL may imply, in
general, lower execution efficiency in comparison to the other related
proposals.

As concerns intensional sets, relatively few (general-purpose)
programming languages provide support for such kind of sets, usually in the
form of list/set comprehension constructs---e.g., SETL
\cite{DBLP:books/daglib/0067831}, Python, Haskell, Miranda \cite{DBLP:journals/eatcs/Turner87},
Scala \cite{Odersky2006}. In all these proposals, list/set comprehension constructs
denote completely specified sets, i.e., sets where all elements have a known
value, and expressions containing list/set comprehensions are always evaluated,
as encountered (though possibly using some form of ``lazy'' evaluation).

Regarding languages for executable specifications, intensional sets are
available, for instance, in ProB \cite{Leuschel00}. ProB is a Prolog program
enabling a range of modeling and verification activities, such as model-based
testing, model checking, etc. Part of its capabilities come from its own solver
but it also uses external solvers such as the Z3 SMT solver, the TLC model
checker and the KodKod constraint solver. In a sense, ProB is similar to
\setlog
\cite{DBLP:journals/jlp/DovierOPR96,DBLP:journals/jar/CristiaR20,DBLP:conf/cade/CristiaR17,setlog},
which is a close relative to JSetL. As we have mentioned, ProB can be used to
check JSetL programs although \setlog would be a more akin option. On the other
hand, both JSetL and \setlog are thought primarily as programming languages
which is not the case of ProB. Besides, some verification activities provided
by ProB can also be performed with JSetL. For example, each solution returned
by a JSetL program can be regarded as a model of that program; and when it
returns no solution, the program is unsatisfiable. In this sense, the JSetL
solver works as a model finder much as ProB for set theoretic formulas.

Also Prolog, as well as extended logic programming languages such as Godel
\cite{DBLP:books/daglib/0095081} and \setlog \cite{DBLP:journals/jlp/DovierOPR96,setlog}, offer
some form of intensionally defined lists/sets (e.g., the built-in predicate
{\fp setof} of Prolog). These facilities are basically based on \emph{set-grouping} \cite{Shmueli1987}, i.e., the ability to collect into an extensional
list/set all the elements satisfying the property characterizing the given
intensional definition. A form of set-grouping is offered also in the first
version of JSetL \cite{Rossi2007} by the method {\fp setof}. Specifically, {\fp
C.setof(x)}, where {\fp x} is an {\fp LVar} and {\fp C} is a {\fp Constraint}
object, returns an {\fp LSet} object whose elements are all possible solutions
for {\fp x} which satisfy {\fp C}.

Though set-grouping works fine in many cases, it may incur in a number of
problems if the formula characterizing the intensional definition contains
unbound variables (other than the control variable) and/or if the set of values
to be collected is not completely determined. Generally speaking, all the above
mentioned proposals lack the ability to perform high-level reasoning on general
formulas involving intensional sets. For instance, these proposals cannot deal
with general formulas like those in Examples \ref{ex:inters} and
\ref{ex:all_positive}.

$\LRIS$ \cite{DBLP:conf/cade/CristiaR17} is a proposal aiming at providing such
capabilities in the context of CLP languages. Using the language of RIS, in
$\LRIS$ we can express very general logic formulas involving intensional sets,
and using the RIS constraint solver we can check their satisfiability and
possibly compute (a finite representation) of all their solutions.

The RIS and the constraints introduced in that context are the same considered
in the current paper. The purposes and methods of that work, however, are quite
different from those considered in JSetL. As a matter of fact, in JSetL  we are
moving within the conventional setting of imperative O-O languages and we are
mainly interested in exploring the potential of using RIS on programming.

As far as we know, this is the first proposal for a conventional programming
language offering support for reasoning about intensional sets. Using RIS and
the other JSetL facilities for constraint solving (including the {\fp setof}
method) we can deal with intensional set definitions in the same way as the
other languages can; but by using RIS constraint solving we can exploit
intensional sets for programming in a more general and original way.

\medskip


Compared to defining a new language, or extending an existing one, where the
desired abstractions are provided as first-class citizens of the language, the
library-based approach adopted in JSetL has the undeniable advantages of $(i)$
being easier to develop; $(ii)$ allowing the library to evolve
independently of the evolution of the language; $(iii)$ having no impact on
the host language, e.g., in terms of execution efficiency; and $(iv)$ not
requiring one to introduce any new formalism, which makes it easy to test the
approach on different languages and communities.

Connected with the last item, our experience with students who have been asked
to learn JSetL (e.g., for thesis work) has shown an easy acceptance of the tool
by them, thanks to the fact that the programming environment is anyway that of
a mainstream language such as Java, although the programming style, at least
for some well-identified parts of the program, can be very different.


\section{Conclusion and future work}\label{conclusion}

In this paper we have presented an extension of the Java library JSetL to
support RIS, and we have shown the usefulness of this extension from the
programming point of view through a number of simple examples. All
sample Java programs shown here are available on-line at the JSetL's home
page \cite{jsetl}.

The advantages of having RIS in JSetL can be summarized as follows:
\begin{enumerate}
\item[$-$] RIS represent a powerful data and control abstraction facility which
integrates and enriches those provided by extensional logical sets, and as such
can be of great help in achieving the goal of making program creation easier
and faster.
\item[$-$] Since intensional sets often play a fundamental role in formal set-based
specifications (e.g., in Z), their presence in JSetL contributes significantly
to the ability to code abstract set-theoretical formulas directly into
(executable) Java+JSetL programs.
%
\item[$-$] The JSetL solver can be used as a theorem-prover for a non-trivial
quantifier-free fragment of set theory; the addition of RIS allows this
fragment to be significantly enlarged, including also sets defined by
properties, as well as (a restricted form of) universally quantified formulas.
\item[$-$] Finally, the fact that RIS are objects, and that partial functions can be
coded as RIS, provides an elegant solution to the problem of dealing with
functions as data in an O-O language; in particular, it is possible to code
also functions as recursive RIS.
\end{enumerate}


As future work, it would certainly be interesting extending the set of atomic
constraints that deal with RIS to the relational operators recently added to
\CLPSET \cite{DBLP:journals/jlp/DovierOPR96,DBLP:journals/jar/CristiaR20}, such
as $\Comp$ for relational composition, $\Inv$ for converse (or inverse), and
$\Id$ for the identity relation over a given set.

This extension could be useful also to improve the possibilities of using JSetL
as a tool for writing executable set-based specifications in Java. Closely
connected with this, a future work could be the precise definition of the
refinement techniques mentioned at the end of Section \ref{practice}. Another
interesting line of work would be to explore how to use JSetL as Eiffel's
contract specification language.

\paragraph{\textbf{Acknowledgments}}
This work has been partially supported by GNCS ``Gruppo Nazionale per il
Calcolo Scientifico''.

\pagebreak
\appendix

\section{Additional formal definitions}\label{app:formal}

\subsection{\label{solved}Solved form formulas}

A $\mathcal{RIS}$-formula $\Phi$ can be seen, without loss of generality, as
$\Phi_\mathcal{S} \land \Phi_\Ur$, where $\Phi_\Ur$ is a $\Ur$-formula, that is
a formula written in the language of the parameter theory of $\LRIS$, and
$\Phi_\mathcal{S}$ is a $\mathcal{RIS}$-formula not containing any
$\Ur$-formula. The $\Phi_\mathcal{S}$ part of $\Phi$ is called a \emph{pure
$\mathcal{RIS}$-formula}.

\begin{definition}[Solved form]\label{def:solved}
Let $\Phi_\mathcal{S}$ be a pure $\mathcal{RIS}$-formula; let $\bar C$, $\bar
D$ and $\bar E$ be either set variables or variable-RIS, $X$ and $Y$ be set
variables but not variable-RIS, $t$ be an $\Ur$-term, and $S$ any set term but
not a RIS. An atom $p$ in $\Phi_\mathcal{S}$ is in \emph{solved form} if it has
one of the following forms:
\begin{enumerate}
\item $\true$

\item\label{i:icfirst} $X = S$ or $X = \{Y | \flt @ \ptt\}$,
and $X$ does not occur in $S$ nor in $\Phi_\mathcal{S} \setminus \{p\}$

\item\label{i:solvedRIS1} $\{X | \flt @ \ptt\} = \varnothing$ or
$\varnothing = \{X | \flt @ \ptt\}$

\item\label{i:solvedRIS2} $\{X | \flt_1 @ \ptt_1\} = \{Y | \flt_2 @ \ptt_2\}$.

\item \label{i:solvedneq} $X \neq S$, and $X$ does not occur in $S$ nor as
the domain of a RIS which is the argument of a $=$ or $\notin$ or $\Cup$
constraint in $\Phi_\mathcal{S}$

\item \label{i:solvednin} $t \notin \bar D$

\item $\Cup(\bar C,\bar D,\bar E)$, and if $\bar C, \bar D \in \Var_{S}$
then $\bar C \not\equiv \bar D$

\item $\bar C \disj \bar D$, and if $\bar C,\bar D \in \Var_{S}$
then $\bar C \not\equiv \bar D$


\end{enumerate}
$\Phi_\mathcal{S}$ is in solved form if all its atoms are in solved form. \qed
\end{definition}

\begin{example}
The following are $\LRIS$ atoms in solved form, occurring in a formula $\Phi$
(where $X$, $D$ and $D_i$ are variables):
\begin{itemize}
\item $X = \{x : D | x \neq 0\}$ and $X$ does not occur
elsewhere in $\Phi$ (note that $X$ and $D$ can be the same variable)
\item $1 \notin \{x : D | x \neq 0\}$
\item $\{x : D_1 | x \mod 2 = 0 @ (x,x)\} = \{x : D_2 | x > 0 @ (x,x+2)\}$
\item $\Cup(X,\riss{D_1}{F}{P},\riss{D_2}{G}{Q})$ and, for any $t$,
there are no constraints $D_1 \neq t$ nor $D_2 \neq t$ in $\Phi$. \qed
\end{itemize}
\end{example}

\subsection{Admissible formulas} \label{ssec:admissible}

We provide sufficient (syntactic) conditions characterizing a sub-language of
$\LRIS$ for which $\SATRIS$ can be proved to terminate and so to be a decision
procedure for that sub-language.

First, we define a transformation $\tau$ of $\mathcal{RIS}$-formulas that
allows us to restrict our attention to a single kind of constraints.

\begin{definition}
Let $\Phi = \Phi_\mathcal{S} \land \Phi_\Ur$ be the input formula, where
$\Phi_\mathcal{S}$ is a pure $\mathcal{RIS}$-formula and $\Phi_\Ur$ is a
$\Ur$-formula; $\Phi_\Ur$ is removed from $\Phi$ and so we only consider its
pure RIS part. Without loss of generality, $\Phi_\mathcal{S}$ can be seen as
$\Phi_1 \lor \dots \lor \Phi_n$, where the $\Phi_i$'s are conjunctions of
primitive $\mathcal{RIS}$-constraints (i.e., all derived constraints have been
replaced by their definitions and the corresponding DNF has been built). Then,
each $\Phi_i$ is transformed into $\Phi_i'$ as follows:
\begin{itemize}
\item constraints of the form $\Cup(A,B,C)$, where $A,B,C$ are either variables
or variable-RIS whose innermost domain variables do not occur elsewhere in
$\Phi_i$, are removed from the formula
\item constraints of the form $A = B$, where neither is $\varnothing$, are
rewritten into $\Cup(A,B,B) \land \Cup(B,A,A)$
\item If one of the arguments of a $\Cup$ constraint is of the form
$\{x_1,\dots, x_n \plus B\}$ then it is replaced by a new variable, $N$, and
$\Cup(\{x_1,\dots, x_n\},B,N)$ is conjoined to the formula
\item all the $\neq$, $\in$, $\notin$, and $\disj$
constraints, and all the remaining $=$ constraints, are removed from the
formula.
\end{itemize}
Hence, $\tau(\Phi) = \Phi_1' \lor \dots \lor \Phi_n'$, where each $\Phi_i'$ is
a conjunction of $\Cup$-constraints. \qed
\end{definition}

\begin{example}\label{ex:tau}\mbox{}
\begin{enumerate}
\item[$(i)$] If $\Phi$ is $\riss{x : D}{F}{(x,y)} \subseteq D \land D \neq \e$ then $\tau(\Phi)$ is
$\Cup(\riss{x : D}{F}{(x,y)},D,D)$

\item[$(ii)$] If $\Phi$ is $\riss{x : D}{F}{(x,y)} \subseteq \risnopattern{x}{A}{G} \land A \subseteq D
\land D \neq \e$ then $\tau(\Phi)$ is $\Cup(\riss{x :
D}{F}{(x,y)},\risnopattern{x}{A}{G},\risnopattern{x}{A}{G}) \land \Cup(A,D,D)$
\qed
\end{enumerate}
\end{example}

The following function allows us to classify set terms occurring as arguments
of $\Cup$-constraints.

\begin{definition}
Let $\type$ be a function that takes a set term $T$ and returns an element in
$\{\stype,\ptype,\utype\}$, where $\stype$ is a constant, $\ptype$
depends on one argument belonging to $\{\stype,\ptype,\utype\}$, and $\utype$
depends on two arguments belonging to $\{\stype,\ptype,\utype\}$. For each
constraint of the form $\Cup(A,B,C)$, the function $\type(T)$ is defined as
follows (note that the definition of $\type$ depends on the position of the
argument in the constraint):
\begin{enumerate}
\item If $T$ is $C$, then:
$\type(C) = \utype(\type(A),\type(B))$
\item If $T$ is either $A$ or $B$, then:
\begin{gather}
\type(\e) = \stype  \\
\type(\{\cdot \plus V\}) = \type(V) \\
\type(\risnopattern{c}{D}{F}) = \type(D)  \\
\type(\riss{c : D}{F}{P}) = \ptype(\type(D)), c \not\equiv P
\end{gather}
and $\type(T)$ remains undefined when $T$ is a variable. \qed
\end{enumerate}
\end{definition}

\begin{example}\label{ex:type_function}
The $\type$ function for the constraint $\Cup(\riss{x : D}{F}{(x,y)},D,D)$ is:
\begin{gather*}
\type(D) = \utype(\type(D),\type(\riss{x : D}{F}{(x,y)})) \\
\iff \type(D) = \utype(\type(D),\ptype(\type(D)))
\end{gather*}
where the computation of $\type$ stops because $D$ is a variable. \qed
\end{example}

\begin{definition}
$\ptype^*(D)$ denotes a $\ptype$ that at some point depends on variable $D$.
\qed
\end{definition}

\begin{definition}[Admissible $\mathcal{RIS}$-formula]\label{d:admissible_formula}
Let $\Phi$ be a $\mathcal{RIS}$-formula not in solved form and $\mathcal{E}$ be
the collection of equalities computed by (recursively) applying the $\type$
function to all the $\Cup$-constraints in $\tau(\Phi)$ and performing all
possible term substitutions. Then $\Phi$ is \emph{non-admissible} iff
$\mathcal{E}$ contains at least one equality of the form $X = \utype(Y,Z)$ such
that:
\begin{itemize}
\item If $X$ depends on $\ptype^*(D)$, for some variable $D$, then $Y$ or $Z$
does not depend on $\ptype^*(D)$; and
\item If $Y$ or $Z$ depends on $\ptype^*(D)$,
for some variable $D$, then $X$ does not depend on $\ptype^*(D)$.
\end{itemize}
All other $\mathcal{RIS}$-formulas are \emph{admissible}. \qed
\end{definition}

\begin{example}
The formula of Example \ref{ex:tau}$(i)$, whose $\type$ function is that of
Example \ref{ex:type_function}, is classified as non-admissible, since $X$
(i.e., $\type(D)$) does depend on $\ptype^*(D)$, while $Z$ (i.e.,
$\ptype(\type(D))$) depends on $\ptype^*(D)$. Conversely, if $\Phi$ is just
$\riss{x : D}{F}{(x,y)} \subseteq D$, that is, $D$ is a variable not occurring
elsewhere in $\Phi$, then $\mathcal{E}$ is empty and $\Phi$ is classified as
admissible. \qed
\end{example}

\begin{example}
Given the formula $\Phi$ of Example \ref{ex:tau}$(ii)$, then the
collection $\mathcal{E}$ for $\Phi$ is
\begin{gather*}
\{\type(\risnopattern{x}{A}{G}) = \utype(\type(\riss{x :
D}{F}{(x,y)}),\type(\risnopattern{x}{A}{G})),
\type(D) = \utype(\type(A),\type(D))\} \\
\iff \\
\{\type(A) = \utype(\ptype(\type(D)),\type(A)),
\type(D) = \utype(\type(A),\type(D))\} \\
\iff \why{substitution} \\
\{\type(A) = \utype(\ptype(\utype(\type(A),\type(D))),\type(A)), \type(D) =
\utype(\type(A),\type(D))\}
\end{gather*}
So $\Phi$ is classified as non-admissible. Given a formula similar to the
previous one, but where the second RIS is a set of pairs, i.e.,
\[
\riss{x : D}{F}{(x,y)} \subseteq \riss{h : A}{G}{(h,w)} \land A \subseteq D
\land D \neq \e
\]
then the final collection $\mathcal{E}$ for this formula is
\[ \{\ptype(\type(A))
= \utype(\ptype(\utype(\type(A),\type(D))),\ptype(\type(A))), \type(D) =
\utype(\type(A),\type(D))\}
\]
so it is classified as admissible. \qed
\end{example}

From the above definitions, it is evident that, if the given formula $\Phi$
does not contain any RIS term, or if all RIS terms possibly occurring in it
have pattern identical to its control term, then $\Phi$ is surely classified as
admissible. Besides, solved form formulas are \emph{admissible}.

Definition \ref{d:admissible_formula} gives only \emph{sufficient} conditions.
In fact, not all formulas classified as non-admissible are indeed formulas that
our solver cannot deal with. Given the formula:
\begin{equation}\label{eq:1}
\riss{x : A}{\false}{(x,y)} \subseteq A \land Y \in A
\end{equation}
any set $A$ satisfying $Y \in A$ is a solution of it. So, we should accept it.
However, according to Definition \ref{d:admissible_formula}, this formula is
classified as non-admissible. Note that the similar formula where the filter is
$\true$ admits only an infinite set solution and is (correctly) classified as
non-admissible.

Accepting or not formula \eqref{eq:1} depends on the satisfiability of the RIS
filter. Checking the satisfiability of the filter, however, cannot be done, in
general, by simple syntactic analysis, i.e., without running the solver on it.
Thus, when aiming at providing a syntactic characterization of admissible
formulas, we must classify formulas disregarding the form of the RIS filters
possibly occurring in them. Finer characterizations would be feasible, however,
considering special forms of RIS filters, such as $\false$ and $\true$. \qed



\section{Pure Java programs}\label{app:java_programs}

\subsection{Printing all permutations of a set of integers---cf. Ex. \ref{ex:permutations}}

\begin{alltt}
    private static int counter = 1;
    private static void swap(Integer[] A,int i,int j) \{
        int aux;
        aux = A[i];
        A[i] = A[j];
        A[j] = aux;
    \}

    private static void allPermutations(Integer[] A,int begin,int n) \{
        int j;
        if(begin == n) \{
            System.out.print(counter + ")  ");
            for(j=0; j<=n; j++)
                System.out.print(A[j] + " ");
            System.out.println();
            counter++;
        \}
        else \{
            for(j = begin;j<=n;j++) \{
                swap(A,begin,j);
                allPermutations(A,begin+1,n);
                swap(A,begin,j);
            \}
        \}
    \}

    public static void allPermutations(Integer[] A) \{
        allPermutations(A,0,A.length-1)
    \}
\end{alltt}

\subsection{Computing the squares of a list of integers---cf. Ex. \ref{ex:mapList}}

\begin{alltt}
    public interface FunctionInt \{
        public int apply(int arg);
    \}

    public static void mapList(Integer[] A,FunctionInt f)  \{
        for(int i=0; i<A.length; i++) \{
            A[i] = f.apply(A[i]);
        \}
    \}

    public static void main (String[] args)  \{
        Integer[] elems = {3,5,7};
        mapList(elems,x -> x*x);
        System.out.println(Arrays.asList(elems));
        \}
\end{alltt}

\section{Additional Java+JSetL programs}\label{app:jsetl_programs}

\subsection{Using JSetL as a theorem prover}\label{min_max}

The following program exploits JSetL to state and prove the property that the
minimum of a set (of integers) is always less or equal than its maximum.
Written in \CLPRIS, the property to be proved is:
\begin{equation*}
\begin{split}
& min \in S \land max \in S \land S \subseteq \{x : S | x \ge min\} \land S
\subseteq \{x : S | x \le max\} \\
& \implies min \leq max
\end{split}
\end{equation*}

To prove this property for all $S$, $min$ and $max$, we prove that the negation
of the above formula:
\begin{equation*}
\begin{split}
& min \in S \land max \in S \land S \subseteq \{x : S | x \ge min\} \land S
\subseteq \{x : S | x \le max\} \\
& \land min > max
\end{split}
\end{equation*}
is false (i.e., the solver raises an exception {\fp Failure}).

\begin{alltt}
    static final int n = 1000000;
    public static void main(String[] args) throws Failure \{
        IntLSet S = new IntLSet();
        IntLVar min = new IntLVar("min",-n,n);
        IntLVar max = new IntLVar("max",-n,n);
        IntLVar x = new IntLVar("x");
        IntLVar y = new IntLVar("y");
        Ris minRis = new Ris(x,S,x.ge(min));
        Ris maxRis = new Ris(x,S,x.le(max));
        Solver solver = new Solver();
        solver.add(min.in(S)
              .and(max.in(S))
              .and(S.subset(minRis))
              .and(S.subset(maxRis))
              .and(min.gt(max)));
        solver.solve();
    \}
\end{alltt}

Note that we need to set a finite domain for the integer variables to allow the
FD solver included in JSetL to work correctly.

\subsection{Ackermann function}

The following class implements the two-argument variant of the original
Ackermann function known as Ackermann-P\'eter function. This implementation
exploits nested recursive RIS. {\fp LList} is the JSetL class that provides
logical lists, i.e., collections of elements of any type similar to logical
sets but where element repetitions and ordering do matter.

\begin{alltt}
public class AckermannPeterFunction \{

    //conventional recursive version
    static int ackermannPeter(int xx, int yy)\{
        if(xx == 0)
            return yy + 1;
        else if(yy == 0)
            return ackermannPeter(xx-1, 1);
        else
            return ackermannPeter(xx-1, ackermannPeter(xx, yy-1));
    \}

    //declarative version using RRIS
    static int risAckermannPeter(int xx, int yy)\{
        IntLVar x = new IntLVar(), y = new IntLVar(), z = new IntLVar();
        IntLVar ze = new IntLVar();
        LList ct = LList.empty().ins(x,y,z);
        LSet domain = new LSet();
        Ris ackermann = new Ris(ct,domain,
           (x.eq(0)
                .and(z.eq(y.sum(1))))
           .or((x.gt(0).and(y.eq(0))
                .and(LList.empty().ins(x.sub(1),1,z).in(domain)))
           .or(x.gt(0).and(y.gt(0))
                .and(LList.empty().ins(x,y.sub(1),ze).in(domain)
                .and(LList.empty().ins(x.sub(1),ze,z).in(domain))))),
        ct,ze);

        IntLVar r = new IntLVar();
        domain.subset(ackermann)
           .and(LList.empty().ins(xx,yy,r).in(ackermann)).check();

        return r.getValue();
    \}

    //sample main method
    public static void main(String[] args)\{
        System.out.println(ackermannPeter(0,5) + " "
           + (ackermannPeter(0,5) == risAckermannPeter(0,5)));
        System.out.println(ackermannPeter(2,0) + " "
           + (ackermannPeter(2,0) == risAckermannPeter(2,0)));
        System.out.println(ackermannPeter(2,1) + " "
           + (ackermannPeter(2,1) == risAckermannPeter(2,1)));
        System.out.println(ackermannPeter(2,2) + " "
           + (ackermannPeter(2,2) == risAckermannPeter(2,2)));
    \}
\}
\end{alltt}

\subsection{Mutually recursive calls}

The following class implements a predicate for checking whether an integer
number is even or not, using mutual recursion.

\begin{alltt}
public class MutuallyRecursiveFunctions \{

    //conventional recursive version
    static boolean isEven(int n)\{
        if(n == 0)
            return true;
        else
            return isOdd(n-1);
    \}
    static boolean isOdd(int n)\{
        if(n == 0)
            return false;
        else
            return isEven(n-1);
    \}

    //declarative version using RIS
    static boolean isEvenRis(int n)\{
        IntLSet e = new IntLSet(), o = new IntLSet();
        IntLVar x1 = new IntLVar(), x2 = new IntLVar();
        Ris even = new Ris(x1,e,x1.eq(0).or(x1.sub(1).in(o)));
        Ris odd = new Ris(x2,o,x2.gt(0).and(x2.sub(1).in(even)));
        return o.subset(odd).and(even.contains(n)).check();
    \}

    //sample main method
    public static void main(String[] args)\{
        for(int i = 0; i < 10; ++i)\{
            System.out.println(i + " " + isEven(i) + " " + isEvenRis(i));
        \}

    \}
\}
\end{alltt}


\begin{thebibliography}{10}
\providecommand \doibase [0]{http://dx.doi.org/}%

\bibitem{Woodcock00}
Woodcock J, Davies J. {\it Using Z: specification, refinement, and proof}.
\newblock Upper Saddle River, NJ, USA: Prentice-Hall, Inc. .
\newblock 1996.

\bibitem{schneider2001b}
Schneider S. {\it The B-method: An Introduction}.
\newblock Cornerstones of computingPalgrave .
\newblock 2001.

\bibitem{DBLP:books/daglib/0067831}
Schwartz JT, Dewar RBK, Dubinsky E, Schonberg E. {\it Programming with Sets -
  An Introduction to {SETL}}.
\newblock Texts and Monographs in Computer ScienceSpringer .
\newblock 1986

\bibitem{DBLP:journals/tplp/CaseauJL02}
Caseau Y, Josset F, Laburthe F. {CLAIRE:} Combining sets, search and rules to
  better express algorithms. {\it Theory Pract. Log. Program.} 2002\string;
  2(6)\string: 769--805.
\newblock \href {\doibase 10.1017/S1471068401001363} {doi:
  10.1017/S1471068401001363}

\bibitem{DBLP:journals/eatcs/Turner87}
Turner D. An overview of Miranda. {\it Bull. {EATCS}} 1987\string; 33\string:
  103--114.

\bibitem{bandicoot}
Cherkashin O, Chrobak J. Bandicoot. 2020.

\bibitem{Abiteboul1991}
Abiteboul S, Grumbach S. A Rule-Based Language with Functions and Sets. {\it
  {ACM} Trans. Database Syst.} 1991\string; 16(1)\string: 1--30.
\newblock \href {\doibase 10.1145/103140.103141} {doi: 10.1145/103140.103141}

\bibitem{Jayaraman1995}
Jayaraman B, Moon K. The SuRE Programming Framework. In:  Alagar VS, Nivat M.
  \kern-2pt, eds. {\it Algebraic Methodology and Software Technology, 4th
  International Conference, {AMAST} '95, Montreal, Canada, July 3-7, 1995,
  Proceedings}. 936 of {\it Lecture Notes in Computer Science}. Springer;
  1995\string: 585

\bibitem{Liu1998}
Liu M. Relationlog: {A} Typed Extension to Datalog with Sets and Tuples. {\it
  J. Log. Program.} 1998\string; 36(3)\string: 271--299.
\newblock \href {\doibase 10.1016/S0743-1066(98)00003-X} {doi:
  10.1016/S0743-1066(98)00003-X}

\bibitem{DBLP:journals/jlp/DovierOPR96}
Dovier A, Omodeo EG, Pontelli E, Rossi G. A Language for Programming in Logic
  with Finite Sets. {\it J. Log. Program.} 1996\string; 28(1)\string: 1--44.
\newblock \href {\doibase 10.1016/0743-1066(95)00147-6} {doi:
  10.1016/0743-1066(95)00147-6}

\bibitem{DBLP:journals/constraints/Gervet97}
Gervet C. Interval Propagation to Reason about Sets: Definition and
  Implementation of a Practical Language. {\it Constraints An Int. J.}
  1997\string; 1(3)\string: 191--244.
\newblock \href {\doibase 10.1007/BF00137870} {doi: 10.1007/BF00137870}

\bibitem{Dovier00}
Dovier A, Piazza C, Pontelli E, Rossi G. Sets and constraint logic programming.
  {\it ACM Trans. Program. Lang. Syst.} 2000\string; 22(5)\string: 861-931.

\bibitem{DBLP:conf/iclp/1993w5}
Omodeo EG, Rossi G. \kern-2pt, eds., {\it Workshop on Logic Programming with
  Sets, in conjunction with {ICLP} 1993, Budapest, Hungary, June 24, 1993};
  1993.

\bibitem{workshop1999}
Jayaraman B, Rossi G. \kern-2pt, eds., {\it Workshop on Declarative Programming
  with Sets};  1999.
\newblock http://people.dmi.unipr.it/gianfranco.rossi/DPS/papers.html.

\bibitem{Rossi2007}
Rossi G, Panegai E, Poleo E. JSetL: a Java library for supporting declarative
  programming in Java. {\it Softw. Pract. Exp.} 2007\string; 37(2)\string:
  115--149.
\newblock \href {\doibase 10.1002/spe.749} {doi: 10.1002/spe.749}

\bibitem{Bergenti2011}
Bergenti F, Chiarabini L, Rossi G. Programming with partially specified
  aggregates in Java. {\it Comput. Lang. Syst. Struct.} 2011\string;
  37(4)\string: 178--192.
\newblock \href {\doibase 10.1016/j.cl.2011.07.002} {doi:
  10.1016/j.cl.2011.07.002}

\bibitem{DBLP:journals/fuin/RossiB15}
Rossi G, Bergenti F. Nondeterministic Programming in {J}ava with {JSetL}. {\it
  Fundam. Inform.} 2015\string; 140(3-4)\string: 393--412.
\newblock \href {\doibase 10.3233/FI-2015-1260} {doi: 10.3233/FI-2015-1260}

\bibitem{DBLP:conf/iclp/DovierPR03}
Dovier A, Pontelli E, Rossi G. Intensional Sets in {CLP}. In:  Palamidessi C.
  \kern-2pt, ed. {\it Logic Programming, 19th International Conference, {ICLP}
  2003, Mumbai, India, December 9-13, 2003, Proceedings}. 2916 of {\it Lecture
  Notes in Computer Science}. Springer; 2003\string: 284--299

\bibitem{DBLP:conf/cade/CristiaR17}
Cristi{\'{a}} M, Rossi G. A Decision Procedure for Restricted Intensional Sets.
  In:  Moura dL. \kern-2pt, ed. {\it Automated Deduction - {CADE} 26 - 26th
  International Conference on Automated Deduction, Gothenburg, Sweden, August
  6-11, 2017, Proceedings}. 10395 of {\it Lecture Notes in Computer Science}.
  Springer; 2017\string: 185--201

\bibitem{Palu:2003:IFD:888251.888272}
{Dal Pal\'{u}} A, Dovier A, Pontelli E, Rossi G. Integrating Finite Domain
  Constraints and {CLP} with Sets. In: PPDP '03. ACM; 2003; New York, NY,
  USA\string: 219--229

\bibitem{Codognet00}
Codognet P, Diaz D. Compiling Constraints in clp({FD}). {\it J. Log. Program.}
  1996\string; 27(3)\string: 185-226.

\bibitem{choco}
{Choco Team} . Choco Solver. https://choco-solver.org/; .

\bibitem{brisset:hal-00938018}
Brisset P, Barnier N. {FaCiLe : a Functional Constraint Library}. In: ; 2001;
  Paphos, Cyprus.

\bibitem{jacop}
Kuchcinski K, Szymanek R. {JaCoP} - {J}ava Constraint Programming solver.
  http://jacop.osolpro.com/; .

\bibitem{gecode}
Tack G, Lagerkvist MZ. Gecode - Generic Constraint Development Environment.
  https://www.gecode.org/; .

\bibitem{Dovier2006}
Dovier A, Pontelli E, Rossi G. Set unification. {\it Theory Pract. Log.
  Program.} 2006\string; 6(6)\string: 645--701.
\newblock \href {\doibase 10.1017/S1471068406002730} {doi:
  10.1017/S1471068406002730}

\bibitem{jsetlman}
Rossi G, Amadini R, Fois A. {\it JSetL User's Manual (Version 3.0)}. .

\bibitem{DBLP:journals/jar/CristiaR20}
Cristi{\'{a}} M, Rossi G. Solving Quantifier-Free First-Order Constraints Over
  Finite Sets and Binary Relations. {\it J. Autom. Reasoning} 2020\string;
  64(2)\string: 295--330.
\newblock \href {\doibase 10.1007/s10817-019-09520-4} {doi:
  10.1007/s10817-019-09520-4}

\bibitem{jsetl}
Rossi G. {JSetL} - A {J}ava Set-oriented Library to support declarative
  (constraint). http://www.clpset.unipr.it/jsetl/; .

\bibitem{setlog}
Rossi G. $\{log\}$.
  http://people.dmi.unipr.it/gianfranco.rossi/setlog.Home.html;  2008.

\bibitem{Leuschel00}
Leuschel M, Butler M. {ProB}: A Model Checker for {B}. In:  Keijiro A, Gnesi S,
  Mandrioli D. \kern-2pt, eds. {\it FME}. 2805 of {\it Lecture Notes in
  Computer Science}. Springer-Verlag; 2003\string: 855--874.

\bibitem{hcvs2017}
Cristi\'a M, Rossi G, Frydman C. Using a Set Constraint Solver for Program
  Verification. In: ; 2017.

\bibitem{DBLP:books/sp/MeyerTouch2009}
Meyer B. {\it Touch of Class: Learning to Program Well with Objects and
  Contracts}.
\newblock Springer .
\newblock 2009

\bibitem{DBLP:journals/cuza/BetarteCLR16}
Betarte G, Campo JD, Luna C, Romano A. Formal Analysis of {A}ndroid's
  Permission-Based Security Model,. {\it Sci. Ann. Comp. Sci.} 2016\string;
  26(1)\string: 27--68.
\newblock \href {\doibase 10.7561/SACS.2016.1.27} {doi: 10.7561/SACS.2016.1.27}

\bibitem{DBLP:journals/cleiej/LunaBCSCG18}
Luna C, Betarte G, Campo JD, Sanz C, Cristi{\'{a}} M, Gorostiaga F. A formal
  approach for the verification of the permission-based security model of
  {A}ndroid. {\it {CLEI} Electron. J.} 2018\string; 21(2).
\newblock \href {\doibase 10.19153/cleiej.21.2.3} {doi: 10.19153/cleiej.21.2.3}

\bibitem{Odersky2006}
Odersky M. The Scala experiment: can we provide better language support for
  component systems?. In:  Morrisett JG, Jones SLP. \kern-2pt, eds. {\it
  Proceedings of the 33rd {ACM} {SIGPLAN-SIGACT} Symposium on Principles of
  Programming Languages, {POPL} 2006, Charleston, South Carolina, USA, January
  11-13, 2006}{ACM}; 2006\string: 166--167

\bibitem{DBLP:books/daglib/0095081}
Hill PM, Lloyd JW. {\it The G{\"o}del programming language}.
\newblock MIT Press .
\newblock 1994.

\bibitem{Shmueli1987}
Shmueli O, Naqvi SA. Set Grouping and Layering in Horn Clause Programs. In:
  Lassez J. \kern-2pt, ed. {\it Logic Programming, Proceedings of the Fourth
  International Conference, Melbourne, Victoria, Australia, May 25-29, 1987
  {(2} Volumes)}{MIT} Press; 1987\string: 152--177.

\end{thebibliography}
\end{document}